\documentclass[aps,prb,twocolumn,preprintnumbers,amsmath,amssymb,superscriptaddress]{revtex4}%

\usepackage{graphicx}%
\usepackage{dcolumn}
\usepackage{amsmath}
\usepackage{color}
\usepackage{ulem}
\usepackage{multirow}

\newcommand{\RN}[1]{\textup{\uppercase\expandafter{\romannumeral#1}}}%

\begin{document}

\title{Charge order breaks time-reversal symmetry in CsV$_3$Sb$_5$}

\author{Rustem Khasanov}
 \email{rustem.khasanov@psi.ch}
 \affiliation{Laboratory for Muon Spin Spectroscopy, Paul Scherrer Institute, CH-5232 Villigen PSI, Switzerland}

\author{Debarchan Das}
 \affiliation{Laboratory for Muon Spin Spectroscopy, Paul Scherrer Institute, CH-5232 Villigen PSI, Switzerland}

\author{Ritu Gupta}
 \affiliation{Laboratory for Muon Spin Spectroscopy, Paul Scherrer Institute, CH-5232 Villigen PSI, Switzerland}

\author{Charles Mielke III}
 \affiliation{Laboratory for Muon Spin Spectroscopy, Paul Scherrer Institute, CH-5232 Villigen PSI, Switzerland}

\author{Matthias Elender}
 \affiliation{Laboratory for Muon Spin Spectroscopy, Paul Scherrer Institute, CH-5232 Villigen PSI, Switzerland}

\author{Qiangwei Yin}
 \affiliation{Department of Physics and Beijing Key Laboratory of Opto-electronic Functional Materials \& Micro-nano Devices, Renmin University of China, Beijing 100872, China}

\author{Zhijun Tu}
 \affiliation{Department of Physics and Beijing Key Laboratory of Opto-electronic Functional Materials \& Micro-nano Devices, Renmin University of China, Beijing 100872, China}

\author{Chunsheng Gong}
 \affiliation{Department of Physics and Beijing Key Laboratory of Opto-electronic Functional Materials \& Micro-nano Devices, Renmin University of China, Beijing 100872, China}

\author{Hechang Lei}
 \email{hlei@ruc.edu.cn}
 \affiliation{Department of Physics and Beijing Key Laboratory of Opto-electronic Functional Materials \& Micro-nano Devices, Renmin University of China, Beijing 100872, China}

\author{Ethan Ritz}
 \affiliation{Department of Chemical Engineering and Materials Science, University of Minnesota, MN 55455, USA}

\author{Rafael M. Fernandes}
 \affiliation{School of Physics and Astronomy, University of Minnesota, Minneapolis, MN 55455, USA}

\author{Turan Birol}
 \affiliation{Department of Chemical Engineering and Materials Science, University of Minnesota, MN 55455, USA}

\author{Zurab Guguchia}
 \email{zurab.guguchia@psi.ch}
 \affiliation{Laboratory for Muon Spin Spectroscopy, Paul Scherrer Institute, CH-5232 Villigen PSI, Switzerland}

\author{Hubertus Luetkens}
 \email{hubertus.luetkens@psi.ch}
 \affiliation{Laboratory for Muon Spin Spectroscopy, Paul Scherrer Institute, CH-5232 Villigen PSI, Switzerland}

\begin{abstract}
The recently discovered vanadium-based kagome metals $A$V$_{3}$Sb$_{5}$ ($A$~=~K,~Rb,~Cs) exhibit superconductivity at low-temperatures and charge density wave (CDW) order at high-temperatures. A prominent feature of the charge ordered state in this family is that it breaks time-reversal symmetry (TRSB), which is connected to the underlying topological nature of the band structure. In this work, a powerful combination of zero-field and high-field muon-spin rotation/relaxation is used to study the signatures of TRSB of the charge order in CsV$_3$Sb$_5$, as well as its anisotropic character. By tracking the temperature evolution of the in-plane and out-of-plane components of the muon-spin polarization, an enhancement of the internal field width sensed by the muon-spin ensemble was observed below $T_{\rm TRSB}=T_{\rm CDW}\simeq95$~K. Additional increase of the internal field width, accompanied by a change of the local field direction at the muon site from the $ab$-plane to the $c$-axis, was detected below $T^\ast\simeq30$~K. Remarkably, this two-step feature becomes well pronounced when a magnetic field of 8~T is applied along the crystallographic $c-$axis, thus indicating a field-induced enhancement of the electronic response at the CDW transition.
These results point to a TRSB in CsV$_3$Sb$_5$ by charge order with an onset of ${\simeq}~95$~K, followed by an enhanced electronic response below ${\simeq}~30$~K. The observed two-step transition is discussed within the framework of different charge-order instabilities, which, in accordance with density functional theory calculations, are nearly degenerate in energy.

\end{abstract}

\maketitle

\section{Introduction}\label{sec:introduction}

The emergence of metallic kagome materials featuring an intricate structural lattice and rich diversity of quantum phases has reinvigorated the quest for finding materials with topological phases built from strongly interacting electrons.\cite{Syozi,TbNature,GuguchiaCSS} This led to the discovery of a vanadium-based kagome metal family $A$V$_{3}$Sb$_{5}$ ($A$~=~K, Rb, Cs),\cite{Ortiz_PRM_2019, Ortiz_PRL_2020,Yin_CPL_2021} which was reported to feature a metallic, topological phase at high temperature, anomalous transverse transport properties including the anomalous Hall effect,\cite{Yang_SciAdv_2020} anomalous Nernst effect,\cite{MHe} unconventional planar Hall effect,\cite{MWang} a transition to a highly tunable unconventional superconducting state at low temperatures,\cite{Yang_arxiv_2021,Zhao_arxiv_2021, Zhang_PRB_2021, Chen_PRL_2021,Wang_PRR_2021,GuguchiaMielke} and time-reversal symmetry breaking (TRSB) charge order. The TRSB isotropic 2$\times$2 charge density wave (CDW) order was first suggested by magnetic-field based scanning tunneling microscopy data \cite{Jiang_NatMat_2021, NShumiya,Wang2021, Zhao_Nature_2021, Liang_PRX_2021, Chen_Nature_2021} and was later widely discussed theoretically.\cite{Denner, TNeupert, Balents, Nandkishore, Feng_ScBull_2021} Several groups observed a reduced CDW symmetry obtained by a 4$\times$1 charge modulation in the CDW state,\cite{Zhiwei_Wang_PRB_2021} which would reduce the rotational symmetry from sixfold $C_6$ to twofold $C_2$.

The combination of zero-field (ZF) and high transverse-field (TF) muon-spin rotation/relaxation ($\mu$SR) experiments has provided direct evidence for TRSB below the onset of charge order in KV$_3$Sb$_5$.\cite{GuguchiaMielke} Similarly, the appearance of spontaneous fields below the charge ordering temperature was also reported for RbV$_3$Sb$_5$.\cite{GuguchiaRVS} ZF-$\mu$SR experiments on the sister compound CsV$_3$Sb$_5$ have reported the onset of the TRSB state at $T_{\rm TRSB}\simeq70$~K,\cite{Yu_arxiv_2021} which is lower than the CDW transition temperature $T_{\rm CDW}\simeq95$~K. A more recent ZF-$\mu$SR study of CsV$_3$Sb$_5$ reported the appearance of spontaneous fields below 50~K.\cite{Smidman} In contrast to ZF-$\mu$SR experiments, Kerr effect measurements reveal the emergence of a TRSB signal in CsV$_3$Sb$_5$ exactly at $T_{\rm CDW}$.\cite{Wu_arxiv_2021} Consequently, the determination of the true onset of spontaneous fields in CsV$_3$Sb$_5$ as well as their in-plane and the out-of-plane anisotropy are of paramount importance as they should intimately relate to the mechanism of charge order.

In this paper, we utilize the combination of ZF and TF-$\mu$SR techniques to probe the $\mu$SR relaxation rates in CsV$_3$Sb$_5$ as a function of temperature, field, and angle $\alpha$ between the in-plane component of the muon-spin polarisation and the crystallographic $a-$axis. The main observation is a two-step increase of the internal field width sensed by the muon-spin ensemble. It consists of a noticeable enhancement at $T_{\rm TRSB}\simeq95$~K, corresponding to the CDW ordering temperature $T_{\rm CDW}$, followed by a stronger increase below $T^\ast\simeq30$~K. An applied magnetic field of 8~T along the crystallographic $c-$axis further enhances the magnetic response below $T_{\rm CDW}$, leading to a much more pronounced two-step increase of the internal field width. Furthermore, the local field at the muon site lies within the $a$-$a-$plane of the crystal in the temperature range from $T_{\rm CDW}$ to $T^\ast$. Below $T^\ast$, the internal field also acquires an out-of-plane component. The absence of the in-plane anisotropy of the internal fields down to $\simeq3$~K was also detected, while the out-of-plane anisotropy remains strong. Our results provide evidence for time-reversal symmetry-breaking in CsV$_3$Sb$_5$ at the onset of charge order, as well as a non-trivial temperature evolution of the electronic response within the charge ordered state. More generally, these results indicate a strong interplay between magnetic and charge channels in this kagome material.

The paper is organized as follows: Section~\ref{sec:Experiment} describes the sample preparation procedure and details of the $\mu$SR experiment. Details of the zero-field and transverse-field $\mu$SR data analysis process are given in Section~\ref{seq:data-analysis}. The results of the zero-field and 8~T transversal-field $\mu$SR experiments are discussed in Section~\ref{sec:results_and discussions}. Conclusions follow in Section~\ref{seq:Conclusions}.

\section{Experimental details} \label{sec:Experiment}

\subsection{Sample preparation and characterization}

Single crystals of CsV$_3$Sb$_5$ were grown following the procedure described in Ref.~\onlinecite{Gupta_Arxiv_2021}. Single crystals with dimensions of $\simeq3\times3\times1$~mm$^3$ were used. As demonstrated in Ref.~\onlinecite{Gupta_Arxiv_2021}, the CsV$_3$Sb$_5$ single crystals possess an obviously hexagonal shape, which allows one to easily distinguish the main crystallographic axes ($a$ and $c$).

The X-ray Laue diffraction images of the studied crystals demonstrate the single crystallinity of the material and confirm the correspondence of the main crystal axes to the sample shape. The Laue images for six individual crystals used in ''in-plane`` rotation experiments are shown in the Supplementary Information.\cite{Supplemental_part}
Based on the results of the Laue studies, the crystals were aligned along $a$ and $c$ axes and mounted on a specially constructed $\mu$SR sample holder (see Sec.~\ref{sec:ZF_muSR} and the Supplementary Information,\cite{Supplemental_part} for further details).

The superconducting transition temperature $T_{\rm c}$ was determined by means of ac susceptibility and was found commonly for all crystals to be $T_{\rm c}\simeq 2.7$~K, in agreement with previously published data.\cite{Ortiz_PRL_2020}
The ac susceptibility curves for two sets of crystals used in zero-field and transverse-field $\mu$SR experiments are presented in the Supplementary Information.\cite{Supplemental_part}

\begin{figure}[htb]
\centering
\includegraphics[width=0.9\linewidth]{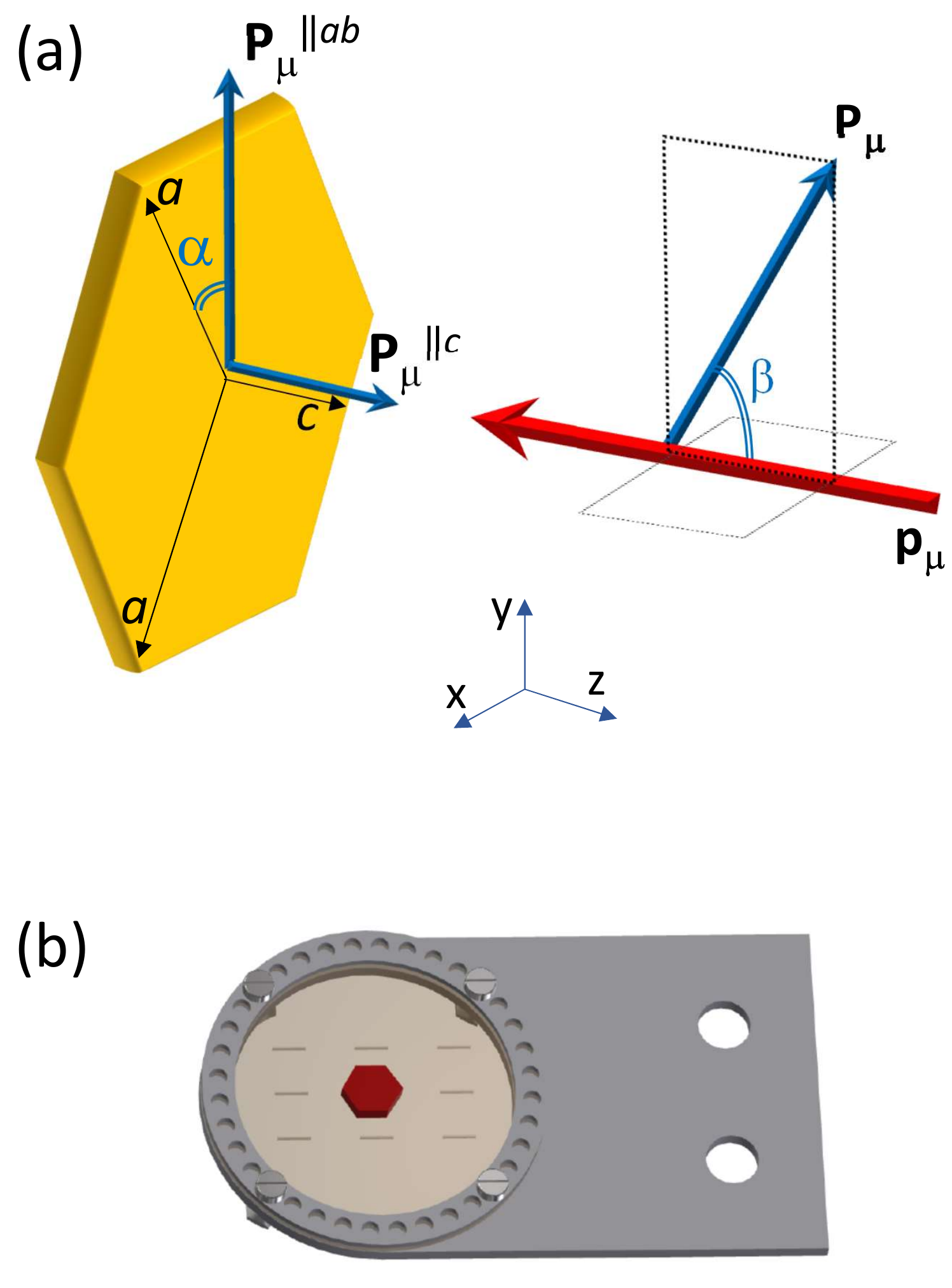}
\caption{(a) A schematic representation of the experimental setup. In two sets of the ZF-$\mu$SR experiments performed at GPS spectrometer, Ref.~\onlinecite{Amato_rsi_2017}, the initial muon-spin polarization ${\bf P}_\mu$ was rotated at an angle $\beta = 60^{\rm o}$ or $5^{\rm o}$  within the vertical ($y-z$) plane. The single-crystalline sample (yellow hexagon) has its $c-$axis aligned along the incoming muon beam (red arrow). The sample can be rotated within the ($x-y$) plane by changing the angle $\alpha$ between the in-plane component of the muon-spin polarisation ${\bf P}_\mu^{\parallel ab}$ and the crystallographic $a-$axis. In experiments performed on the HAL-9500 spectrometer, the initial muon-spin polarization was aligned perpendicular to the muon-momentum ($\beta\simeq90^{\rm o}$). The external magnetic field $B_{\rm ext} =8$~T was applied parallel to the muon-momentum (${\bf B}_{\rm ext} \parallel {\bf p}_\mu$) and parallel to the crystallographic $c-$axis (${\bf B}_{\rm ext} \parallel c$). (b) The sample holder for ''in-plane`` rotation experiments. The angle $\alpha$ can be changed with $10^{\rm o}$ step (see text for details).}
\label{fig:Experimental_Setup}
\end{figure}

\begin{figure*}[htb]
\centering
\includegraphics[width=0.9\linewidth]{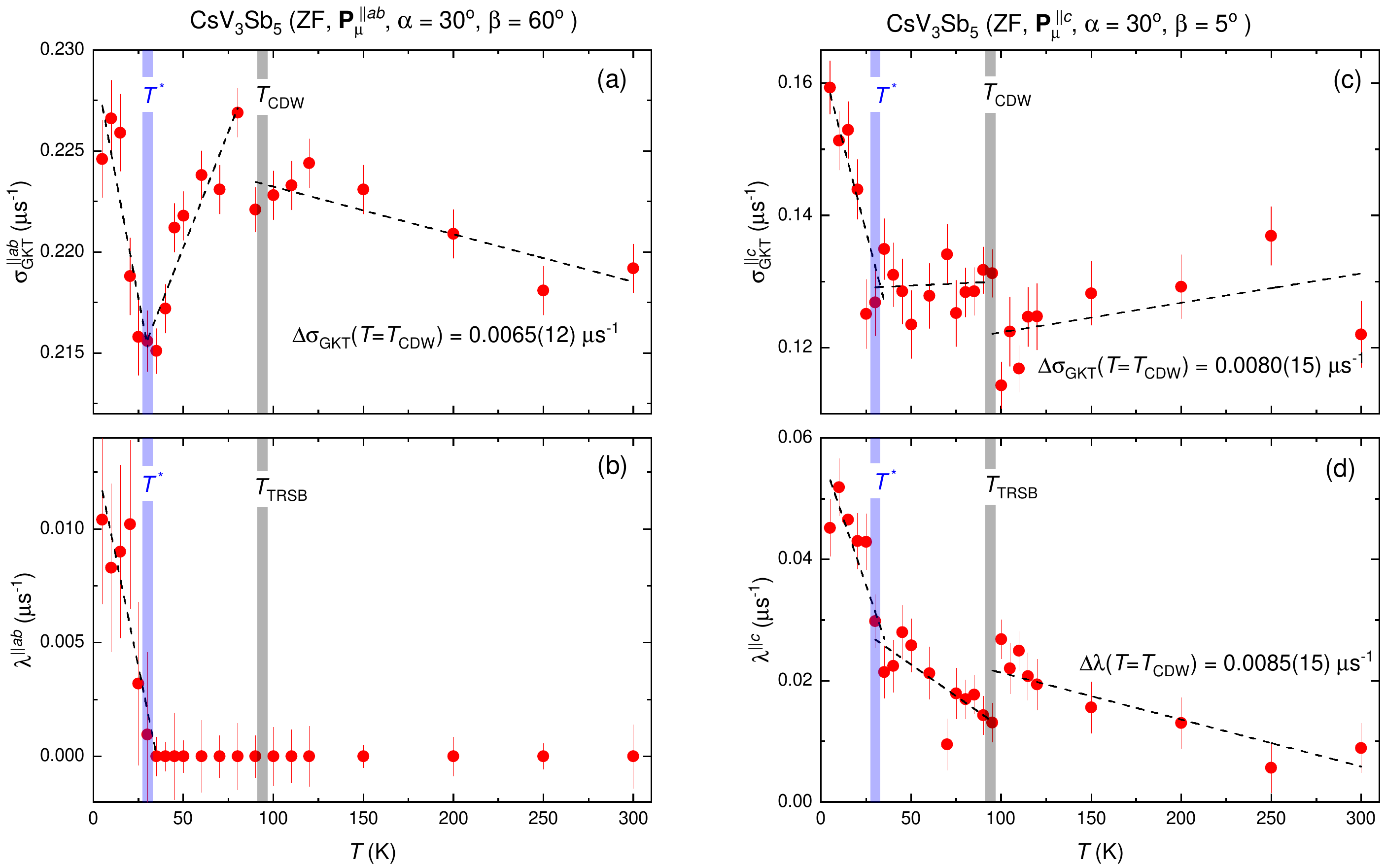}
\caption{(a)--(b) Temperature dependencies of the Gaussian Kubo-Toyabe [$\sigma_{\rm GKT}$, panel (a)] and exponential [$\lambda$, panel (b)] relaxation rates obtained from fits of the ${\bf P}^{\parallel ab}_\mu(t)$ components of the muon-spin polarization in ZF-$\mu$SR experiments. The angles $\alpha$ and $\beta$ were set to $30^{\rm o}$ and $60^{\rm o}$, respectively. (c)--(d) The same as in panels (a) and (b), but for ${\bf P}^{\parallel c}_\mu(t)$ and $\beta=5^{\rm o}$. The dashed lines are linear fits in the temperature regions: $5{\rm ~K} \leq T \leq 30$~K, $30{\rm ~K} \leq T \leq 95$~K, and $95{\rm ~K} \leq T \leq 300$~K. The broad gray and violet lines in panels (a) and (c) represent the CDW ordering temperature $T_{\rm CDW}$ and the characteristic temperature $T^\ast$. The broad lines in panels (b) and (d) represent the TRSB transition temperature $T_{\rm TRSB}$ and $T^\ast$, respectively. }
\label{fig:T-scan}
\end{figure*}

\

\subsection{Muon spin rotation/relaxation experiments}\label{sec:muSR_technique}

The muon spin rotation/relaxation ($\mu$SR) experiments were carried out at the $\pi$M3 and $\pi$E3 beam lines by using the GPS, Ref.~\onlinecite{Amato_rsi_2017}, and HAL-9500 spectrometers (Paul Scherrer Institute, Switzerland). The $\mu$SR measurements were performed at temperatures ranging from $\simeq 3$ to 300~K. The 100\% spin-polarized ''surface`` muons with a momentum of $p_\mu\simeq28.6$~MeV/c were implanted into the CsV$_3$Sb$_5$ crystals along the $c-$axis [see Fig.~\ref{fig:Experimental_Setup}~(a)]. Muons thermalize rapidly without a significant loss of their initial spin polarization and stop in matter at the depth of about 0.15~g/cm$^2$. For CsV$_3$Sb$_5$, with the density of $\sim 5$~g/cm$^3$, this corresponds to a depth of $\simeq0.3$~mm. With the CsV$_3$Sb$_5$ crystal thickness of $\simeq1$~mm, all ''surface`` muons stop in the sample, so the use of degraders, as it is required in $\mu$SR studies of thin single-crystal samples,\cite{Khasanov_FeSe-Intercalated_PRB_2016} was not necessary.

\subsection{ZF-$\mu$SR experiments}\label{sec:ZF_muSR}

Experiments with zero applied field were performed at the GPS spectrometer. Measurements were made by varying two angles: $\beta$, the angle between the initial muon-spin polarization ${\bf P}_\mu$ and the muon momentum ${\bf p}_\mu$; and $\alpha$, the angle between the crystal's $a-$axis and the in-plane component of the muon-spin polarization ${\bf P}^{\parallel ab}_\mu$ [see Fig.~\ref{fig:Experimental_Setup}~(a)].
Both ''in-plane`` and ''out-of-plane`` ZF-$\mu$SR experiments were conducted. In the ''in-plane`` rotation experiments, the angle $\beta$ was kept at the maximum value allowed by the spin-rotator setup at the $\pi$M3 beam-line ($\beta\simeq 60^{\rm o}$, Ref.~\onlinecite{Amato_rsi_2017}). Temperature scans were performed for three different values of $\alpha=0^{\rm o}$, $30^{\rm o}$ and $60^{\rm o}$. In these experiments, the time evolution of ${\bf P}^{\parallel ab}_\mu$  component of the muon-spin polarization [Fig.~\ref{fig:Experimental_Setup}~(a)] was accessed.
The ''out-of-plane`` experiments were conducted with $\alpha$ and $\beta$ set to $30^{\rm o}$ and $5^{\rm o}$, respectively. The temperature evolution of ${\bf P}^{\parallel c}_\mu$ component of the muon-spin polarization was measured. Note that $\beta=5^{\rm o}$ corresponds to the smallest possible muon-spin rotated angle at the GPS instrument.\cite{Amato_rsi_2017}

In order to vary the angle $\alpha$, a special sample holder was constructed [Fig.~\ref{fig:Experimental_Setup}~(b)]. It consists of an aluminum support plate and two sample mounting rings. Each sample ring has 36 holes, which allows for rotation with a $10^{\rm o}$ step relative to the support plate. The $a-$ and $c-$axis-aligned CsV$_3$Sb$_5$ crystals were glued on a 25~$\mu$m aluminum foil, which was further attached to one of the aluminum sample rings. The second ring was covered by a 25~$\mu$m thin layer of Kapton in order to prevent the crystals from falling down inside the cryostat. Note that the Kapton and aluminum foils are fully transparent for the ''surface`` muons, which allows us to use the advantage of the so-called ''Veto`` mode. The ''Veto`` mode rejects the muons missing the sample and, as a consequence, reduces the background of the $\mu$SR response to nearly zero (see Ref.~\onlinecite{Amato_rsi_2017}, for a detailed explanation of the ''Veto`` mode principle). Photos of six CsV$_3$Sb$_5$ crystals mounted on the muon sample holder and the background estimate are given in the Supplementary Information.\cite{Supplemental_part}

\subsection{TF-$\mu$SR experiments}

Experiments with the magnetic field applied transversal to the initial muon-spin polarisation (TF-$\mu$SR experiments) were performed at the HAL-9500 spectrometer. The initial muon-spin polarization was set perpendicular to the muon-momentum [$\beta\simeq90^{\rm o}$, Fig.~\ref{fig:Experimental_Setup}~(a)]. The external magnetic field, $B_{\rm ext} =8$~T, was applied parallel to the muon-momentum (${\bf B}_{\rm ext} \parallel {\bf p}_\mu$) {\it i.e.} parallel to the crystallographic $c-$axis (${\bf B}_{\rm ext} \parallel c$). In these experiments the time evolution of the ${\bf P}^{\parallel ab}_\mu$  component of the muon-spin polarization [Fig.~\ref{fig:Experimental_Setup}~(a)] was accessed.
Crystals were mounted on the sample-holder made of 99.999\% pure silver, and can bee seen in a photo shown in the Supplementary Information.\cite{Supplemental_part}

\subsection{$\mu$SR data analysis procedure}\label{seq:data-analysis}

The zero-field $\mu$SR spectra were fitted using the Gaussian Kubo-Toyabe (GKT) relaxation function,\cite{Hayano_PRB_1979,MUSRFIT} describing the nuclear moment response, multiplied by an additional exponential term:
\begin{equation}
A_i^{\rm ZF}(t) = A_{0,i}^{\rm ZF}\left[ \frac{1}{3}+\frac{2}{3}(1-\sigma_{\rm GKT}^2t^2)\; e^{-\sigma_{\rm GKT}^2 t^2/2}\right]\;e^{-\lambda t}.
 \label{eq:P_sample}
\end{equation}
Here, $A_{0,i}^{\rm ZF}$ is the initial asymmetry of the $i-$th positron detector at $t=0$, $\sigma_{\rm GKT}$ is the GKT relaxation rate, and $\lambda$ is the exponential relaxation rate. Note that Eq.~\ref{eq:P_sample} is widely used to analyze the ZF-$\mu$SR data in most TRSB $\mu$SR studies.\cite{GuguchiaMielke, Luke_Nature_1998, Hillier_PRL_2012, Biswas_PRB_2013, Grinenko_SRO_NatPhys_2020}

The TF-$\mu$SR data were analyzed as:
\begin{equation}
A_i^{\rm TF}(t) = A_{0,i}^{\rm TF}\; \cos(\gamma_\mu B_{\rm int}t+\phi_i) \; e^{-\sigma^2 t^2/2}\; e^{-\lambda t}.
 \label{eq:P_sample_TF}
\end{equation}
Here $B_{\rm int}$ is the internal field at the muon site, $\phi_i$ is the initial phase of the muon-spin ensemble, $\gamma_\mu=2\pi\times~135.5$~MHz/T is the muon gyromagnetic ratio, and $\sigma$ is the Gaussian relaxation rate.

In the above Eqs.~\ref{eq:P_sample} and \ref{eq:P_sample_TF}, $\sigma_{\rm GKT}$ and $\sigma$ relaxation rates mainly account for the nuclear moment contribution, which is assumed to be static  within the $\mu$SR time window. As  discussed previously for KV$_{3}$Sb$_{5}$  and RbV$_{3}$Sb$_{5}$, the exponential relaxation rate $\lambda$ is mostly sensitive to the temperature dependence of the electronic contribution to the muon spin relaxation.\cite{GuguchiaMielke, GuguchiaRVS} One cannot exclude, however, subtle effects owing to changes in the electric field gradients in the charge ordered state.\cite{Sonier_PRB_2002} The details of the ZF- and TF-$\mu$SR data analysis procedure are discussed in the Supplementary Information.\cite{Supplemental_part}

\begin{figure}[htb]
\centering
\includegraphics[width=1.0\linewidth]{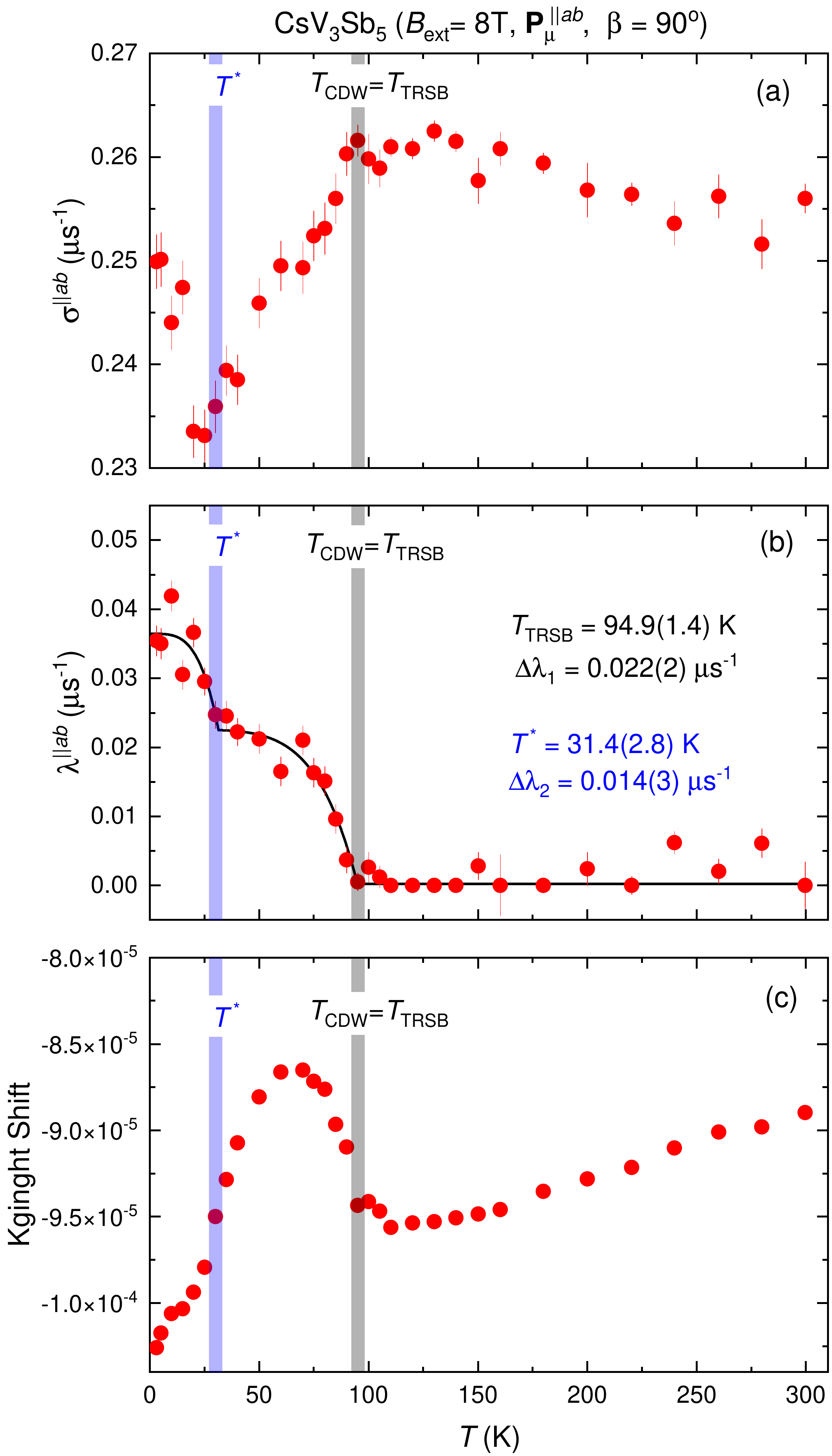}
\caption{Temperature evolution of the Gaussian relaxation rate [$\sigma$, panel (a)], the exponential relaxation rate [$\lambda$, panel(b)], and the  Knight Shift [$(B_{\rm int}-B_{\rm ext})/B_{\rm ext}$, panel (c)] obtained in TF-$\mu$SR experiments. An external magnetic field $B_{\rm ext}=8$~T was applied perpendicular to the initial muon-spin polarization and parallel to the crystallographic $c-$axis of the CsV$_3$Sb$_5$ sample.
 }
\label{fig:TF_HAL}
\end{figure}

\section{Experimental results and discussions}\label{sec:results_and discussions}

The parameters obtained from the analysis of ZF- and TF-$\mu$SR data are summarized in Figs.~\ref{fig:T-scan}, \ref{fig:TF_HAL}, and ~\ref{fig:In-plane_rotation}.

Figure~\ref{fig:T-scan} compares the results of the ''in-plane`` and ''out-of-plane``  ZF-$\mu$SR experiments conducted at $\alpha$~=~30$^{\rm o}$. The left and right columns of Fig.~\ref{fig:T-scan} represent the parameters obtained from the fits of the ${\bf P}^{\parallel ab}_\mu(t)$ and ${\bf P}^{\parallel c}_\mu(t)$ components of the muon-spin polarization, respectively.

Figure~\ref{fig:TF_HAL} shows the temperature evolution of the fit parameters obtained from the 8~T TF-$\mu$SR experiments. Panels (a), (b), and (c) refer to the temperature dependencies of $\sigma$, $\lambda$, and the experimental muon Knight shift $(B_{\rm int}-B_{\rm ext})/B_{\rm ext}$, respectively.

Figure~\ref{fig:In-plane_rotation} shows the temperature dependencies of the Gaussian Kubo-Toyabe [panel (a)] and the exponential [panel (b)] relaxation rates obtained in the ''in-plane`` rotation ZF-$\mu$SR set of experiments. The time evolution of the in-plane component of the muon-spin polarization ${\bf P}^{\parallel ab}_\mu(t)$ was analyzed. The closed circles, open circles and closed triangles depict the data taken at $\alpha=0^{\rm o}$, $30^{\rm o}$, and $60^{\rm o}$, respectively.

The experimental results presented in Figs.~\ref{fig:T-scan}, \ref{fig:TF_HAL} and ~\ref{fig:In-plane_rotation} are closely related to each other and display distinct features, which will be discussed in the following Sections.

\subsection{Two-step feature and out-of-plane anisotropy of the internal field}

Figure~\ref{fig:T-scan} shows that three of four ZF relaxation rates, namely $\lambda$ in the ${\bf P}^{\parallel c}_\mu$ set of experiments ($\lambda^{\parallel c}$) and the two Kubo-Toyabe relaxation rates 
from the ${\bf P}^{\parallel ab}_\mu$ and ${\bf P}^{\parallel c}_\mu$ sets of data ($\sigma_{\rm GKT}^{\parallel ab}$ and $\sigma_{\rm GKT}^{\parallel c}$) display a sudden change across $T_{\rm CDW}\simeq 95$~K (indicated by the thick grey lines in all four panels of Fig.~\ref{fig:T-scan}). Namely, both $\sigma_{\rm GKT}$ suddenly increase, while $\lambda$ decreases when crossing the CDW transition temperature. The absolute value of the jump, as estimated from the linear fits of the relaxation rate data above and below $T_{\rm CDW}$, was found to be the same (within the experimental uncertainty) and corresponds to $\simeq 0.0075(20)$~$\mu {\rm s}^{-1}$ [see Figs.~\ref{fig:T-scan} (a), (c), and (d)]. The appearance of a step-like change of both $\sigma_{\rm GKT}$ and $\lambda$ agrees with the first-order nature of the CDW transition of CsV$_3$Sb$_5$ reported in Refs.~\onlinecite{Ortiz_PRM_2019, Ortiz_PRL_2020, Yin_CPL_2021, Zhou_PRB_2021, Jiang_NatMat_2021, Zhao_Nature_2021, Liang_PRX_2021, Chen_Nature_2021, Tan_PRX_2021, Li_PRX_2021, Wang_PRB_2021}

Upon lowering the temperature below $T_{\rm CDW}$, two different slopes above and below the characteristic temperature $T^\ast\simeq30$~K are detected for $\lambda^{\parallel c}(T)$ [Fig.~\ref{fig:T-scan}~(d)]. At the same time, $\lambda^{\parallel ab}(T)$ remains at zero for $T\gtrsim T^\ast$ and increases with decreasing temperature below $T^\ast$ [Fig.~\ref{fig:T-scan}~(b)].
In a simplified description, the finite exponential relaxation rate observed in the ${\bf P}_\mu^{\parallel c}$ experiments accounts for the width of the field distribution in the $ab-$plane ($\Delta B^{\parallel}$), while a finite relaxation along ${\bf P}_\mu^{\parallel ab}$  is associated with the field components from the $ac-$ and/or $bc-$planes ($\Delta B^\perp$). Accounting for the experimental data presented in Figs.~\ref{fig:T-scan}~(b), \ref{fig:T-scan}~(d), and \ref{fig:In-plane_rotation}~(b), this implies that for $T^\ast\lesssim T \lesssim T_{\rm CDW}$ the internal field on the muon position has only $B^\parallel$ components, while for $T \lesssim T^\ast$ both $B^\parallel$ and $B^\perp$ components are present.

The increase of the exponential contribution to the internal field width is also accompanied by a non-monotonic temperature dependence of the Gaussian contribution to the internal field width. Below $T_{\rm CDW}\simeq 95$~K, the Gaussian relaxation rate changes in two steps. In the region $T^\ast \lesssim T \lesssim T_{\rm CDW}$,  $\sigma_{\rm GKT}^{\parallel ab}(T)$ decreases, while $\sigma_{\rm GKT}^{\parallel c}(T)$ stays nearly constant with decreasing temperature. Below $T^\ast\simeq30$~K, the temperature dependence of both relaxation components change slope and begin to increase.

While the increase of the exponential relaxation $\lambda^{\parallel c}(T)$ of CsV$_{3}$Sb$_{5}$ is consistent with the onset of time-reversal symmetry-breaking at $T_{\rm CDW}$ ($T_{\rm TRSB}\simeq T_{\rm CDW}$), high-field ${\mu}$SR experiments are essential to confirm this effect,  as the ZF and weak TF-$\mu$SR data can be subtly affected by the onset of different charge orders even without the presence of TRSB. Following Ref.~\onlinecite{GuguchiaMielke}, the application of a high magnetic field leads to a strong enhancement of the electronic contribution to the relaxation rate. Comparison of the relaxation rates obtained in ZF- and TF-$\mu$SR experiments confirms that this is actually the case. Indeed, $\sigma^{\parallel ab}(T)$ measured at $B_{\rm ext}=8$~T [Fig.~\ref{fig:TF_HAL}~(a)] reproduces the temperature evolution of $\sigma_{\rm GKT}^{\parallel ab}$ collected in ZF-$\mu$SR studies [Fig.~\ref{fig:T-scan}~(a)]. The absolute values of both $\sigma^{\parallel ab}$ and $\sigma_{\rm GKT}^{\parallel ab}$ stay almost the same within 10-15\%. In contrast, the exponential component is strongly affected by the applied field. There is a factor of $\sim1.5-2$ enhancement of $\lambda^{\parallel ab}(T)$ in the $T\lesssim30$~K region and a full recovery of the $\lambda^{\parallel ab}$ component for temperatures between $30{\rm ~K}\lesssim T \lesssim 95$~K. This gives rise to a well-pronounced two-step feature at high fields. Note that such a two-step increase of the relaxation rate is also observed in the sister compound RbV$_{3}$Sb$_{5}$.\cite{GuguchiaRVS}

To estimate the onset temperatures associated with this two-step transition, the TF $\lambda^{\parallel ab}(T)$ was fitted with a double-stage power-law functional form. While this is not a microscopically derived function, it allows us to gain quantitative insight about the temperatures involved. The fit function assumes that each state is characterised by its own transition temperature ($T_{\rm TRSB}$ and $T^\ast$), the enhancement of $\lambda(T)$ due to spontaneous magnetic fields ($\Delta\lambda_1$ and $\Delta\lambda_2$), and the power-law exponent  ($n_1$ and $n_2$):

\begin{widetext}
\begin{equation}
   \lambda(T) =
\begin{cases}
0  , & T>T_{\rm TRSB} \\
\Delta\lambda_1\left[1-\left(\frac{T}{T_{\rm TRSB}}\right)^{n_1}\right] , & T^\ast<T<T_{\rm TRSB}\\
\Delta\lambda_1\left[1-\left(\frac{T}{T_{\rm TRSB}}\right)^{n_1}\right]
+\Delta\lambda_2\left[1-\left(\frac{T}{T^\ast}\right)^{n_2}\right], &T<T^\ast.
\end{cases}.
 \label{eq:lambda}
\end{equation}
\end{widetext}
The solid line in Fig.~\ref{fig:TF_HAL}~(b) represents the results of the fit. The parameters are: $T_{\rm TRSB}=94.9(1.4)$~K, $\Delta\lambda_1=0.0022(2)$~$\mu{\rm s}^{-1}$, $n_1=5.3(1.5)$ and  $T^\ast=31.4(2.8)$~K, $\Delta\lambda_2=0.0014(3)$~$\mu{\rm s}^{-1}$, $n_2=3.7(2.5)$ for the first and the second step, respectively.

\begin{table*}[htb]
     \centering
     \caption{The results of the density functional theory (DFT) calculations for CsV$_3$Sb$_5$ obtained in Ref.~\onlinecite{Gupta_CsV3Sb-pressure_unpublished}. The columns ''Space Symmetry``, ''Lattice Symmetry``, ''Rotation Symmetry``, and ''$E_{\rm CDW}-E_{\rm parent}$ per f.u.`` denote the space group of the CDW unit cell, the rotation symmetry along the $c-$axis, and the internal energy difference between the CDW ordered and the parent (non-ordered) states per formula unit, respectively.\\ }
     \begin{tabular}{c|ccccc}
 \hline \hline
CDW order    &  Space group &Rotation symmetry&$E_{\rm CDW}-E_{\rm parent}$ per f.u.\\
 \hline
Planar tri-hexagonal&$P6/mmm$ (\#191)&$C_6$ (6-fold)& $-21$~meV\\
Superimposed tri-hexagonal Star-of-David&$P6/mmm$ (\#191)&$C_6$ (6-fold)& $-17$~meV\\
Staggered tri-hexagonal&$Fmmm$ (\#69)&$C_2$ (2-fold)& $-24$~meV\\
 \hline  \hline
     \end{tabular}
     \label{tab:DFT-results}
 \end{table*}

The combination of ZF-${\mu}$SR and high-field TF-${\mu}$SR results on CsV$_{3}$Sb$_{5}$ provides an indication of time-reversal symmetry-breaking below the onset of charge order $T_{\rm TRSB}=T_{\rm CDW}\simeq95$~K. This agrees well with the previous reports on KV$_{3}$Sb$_{5}$ \cite{GuguchiaMielke} and RbV$_{3}$Sb$_{5}$,\cite{GuguchiaRVS} indicating that the TRSB effect is strongly connected to the charge-density wave transition for all three members of this kagome metal family. It has to be mentioned, however, that the effects of TRSB in CsV$_{3}$Sb$_{5}$ are much less pronounced than for the sister compounds KV$_{3}$Sb$_{5}$\cite{GuguchiaMielke} and RbV$_{3}$Sb$_{5}$\cite{GuguchiaRVS} and might easily be overlooked by less precise measurements. It seems to be especially vital that high quality single crystals are used for the investigation.

One way to understand these results is that the internal fields experienced by the muons are generated by orbital currents associated with a complex CDW order parameter.\cite{Denner,Balents,Nandkishore,Haldane,Varma} Within this framework, muons can couple to the fields generated by these loop currents, resulting in an enhanced internal field width sensed by the muon-spin ensemble.
A direct connection between the orbital current patterns and the observed internal fields remains a challenge.
The first attempt in calculating possible field directions was made in Ref.~\onlinecite{Yu_arxiv_2021} by considering a few possible orbital current configurations which are allowed by symmetry.\cite{Feng_PRB_2021} At the moment, it is difficult to proceed deeper into the subject. Further theoretical studies, including the exact determination of the muon-stopping site(s) and possible configurations of the  orbital currents are needed.

The increase of the exponential relaxation rates below the characteristic temperature $T^\ast$ ${\simeq}$~30~K is suggestive of another transition that modifies the loop currents formed at $T_{\rm TRSB}=T_{\rm CDW}\simeq95$~K. In Ref.~\onlinecite{Yu_arxiv_2021} the low-temperature increase was interpreted as a change in symmetry of the orbital currents within the same chiral flux state. However, there are experimental indications that some kagome metals (including CsV$_{3}$Sb$_{5}$) may exhibit two charge-order transitions.\cite{MShi,ZhengNMR} For instance, the coexistence of the tri-hexagonal and Star-of-David CDW patterns in CsV$_{3}$Sb$_{5}$ was reported by ARPES.\cite{MShi} Similarly, NMR/NQR experiments point to the Star-of-David CDW at high temperatures, followed by an additional charge modulation below ${\sim}$~40~K.\cite{ZhengNMR} Breaking of the sixfold symmetry of the CDW state was reported experimentally for CsV$_{3}$Sb$_{5}$.\cite{Zhao_Nature_2021, LNie,Xiang} More broadly, the pressure-dependent $\mu$SR data on the RbV$_3$Sb$_5$ compound  also indicates two CDW transitions.\cite{GuguchiaRVS}

Our previous low-temperature, pressure-dependent $\mu$SR data on CsV$_3$Sb$_5$ revealed a strong change in the superfluid density within the CDW phase as pressure was varied.\cite{Gupta_CsV3Sb-pressure_unpublished} Combined with the first-principles calculations, this was interpreted as indicative of a change in the zero-temperature CDW ground state as a function of pressure. Such a behavior might also be consistent with a change in the CDW state as a function of temperature for zero applied pressure. Table~\ref{tab:DFT-results} reports our density functional theory (DFT) results from Ref.~\onlinecite{Gupta_CsV3Sb-pressure_unpublished} for the relative energies of three of the possible CDW states at ambient pressure -- namely, the planar tri-hexagonal, staggered tri-hexagonal, and superimposed tri-hexagonal Star-of-David (see Ref.~\onlinecite{Christensen_arxiv_2021} for the schematics of each CDW configuration). As follows from Table~\ref{tab:DFT-results}, the energy differences between these states are so small (about 5~meV) that the issue of which CDW state has the lowest energy is likely to be affected by finite-temperature effects.\cite{Ratcliff_PRM_2021, Subedi_PRM_2022} Such effects might be associated with either the ''electronic temperature`` within DFT or the entropy contribution to the free-energy, which is not captured by DFT.\cite{Christensen_arxiv_2021} Moreover, phonon modes associated with other CDW states with additional modulation along the $z-$axis not considered here ({\it e.g.} a 2$\times2\times$4 configuration) are also unstable, expanding the landscape of possible CDW configurations even further.\cite{Wu_arxiv_2022}

It is important to note, however, that these non-spin-polarized first-principles calculations that do not take spin-orbit coupling into account refer only to the real component of the complex CDW order parameter. As such, they do not capture the role of orbital currents. Furthermore, our $\mu$SR results show that not only time-reversal symmetry is broken below $T_{\rm CDW}$, but that the internal magnetic fields rotate out of the plane below $T^\ast$. Therefore, a full understanding of the two-step transition observed here will require a more in-depth analysis of the role of the imaginary component of the complex CDW order parameter. While first-principles calculations of orbital currents are challenging and likely cannot be captured at the generalized gradient approximation level, phenomenological approaches reveal an interesting connection between the real and imaginary components,\cite{Balents, Nandkishore} which deserve further exploration.

\subsection{Absence of the in-plane anisotropy of the internal field width}

\begin{figure}[htb]
\centering
\includegraphics[width=0.9\linewidth]{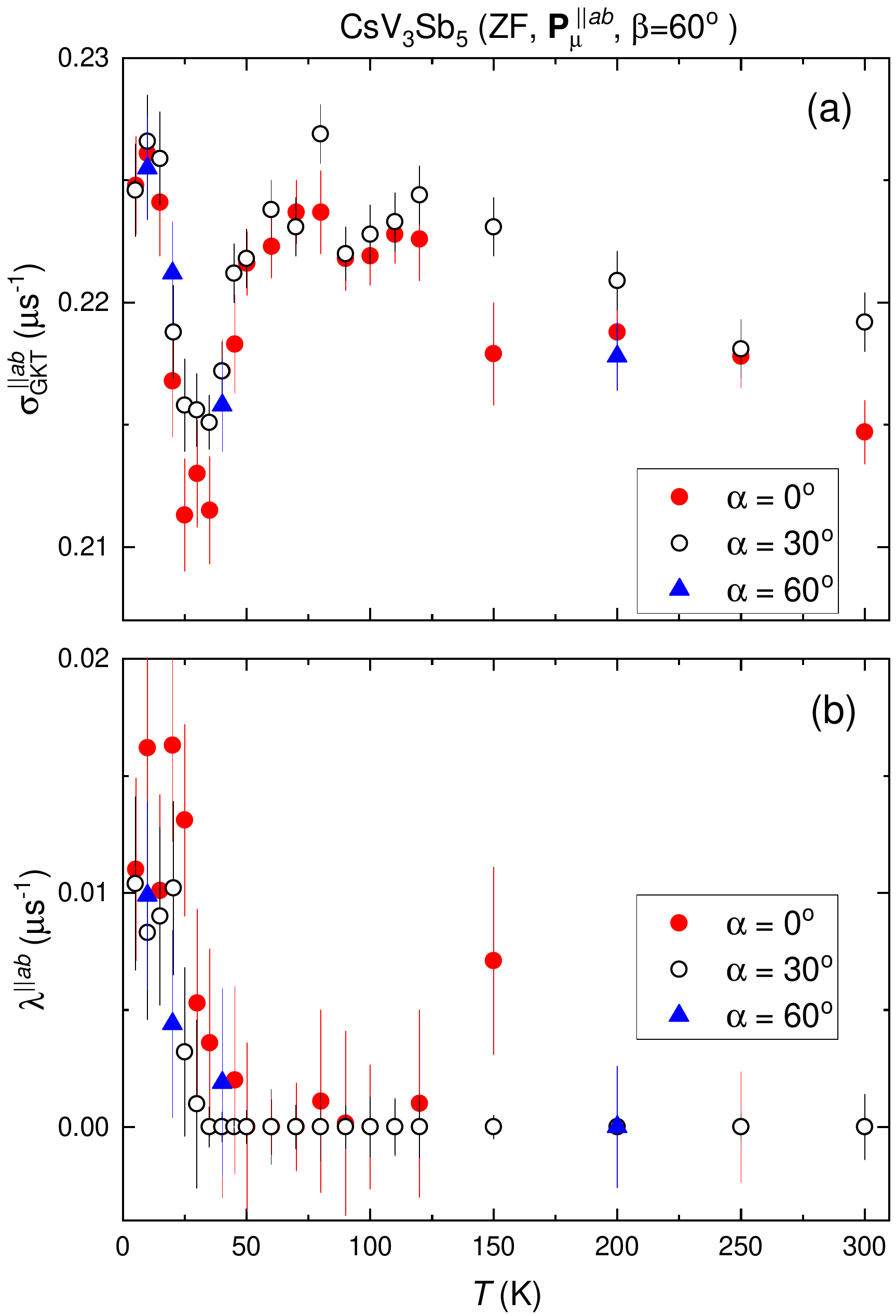}
\caption{Temperature dependencies of the Gaussian Kubo-Toyabe [$\sigma_{\rm GKT}$, panel (a)] and exponential [$\lambda$, panel (b)] relaxation rates obtained in ''in-plane`` rotation ZF-$\mu$SR experiments for three values of the angle $\alpha=0^{\rm o}$, $30^{\rm o}$, and $60^{\rm o}$. }
\label{fig:In-plane_rotation}
\end{figure}

Figures~\ref{fig:In-plane_rotation}(a) and (b) show the temperature dependencies of the Gaussian Kubo-Toyabe and the exponential relaxation rates, respectively, obtained in the ''in-plane`` rotation ZF-$\mu$SR experiments for three values of the angle $\alpha=0^{\rm o}$, $30^{\rm o}$, and $60^{\rm o}$ between the crystal's $c-$axis and the in-plane component of the muon-spin polarization ${\bf P}_{\mu}^{\parallel ab}$ (see Fig.~\ref{fig:Experimental_Setup}). The temperature dependencies of both relaxation rates
($\sigma_{\rm GKT}$ and $\lambda$) do not depend on the angle $\alpha$ within experimental accuracy. Thus, the internal field width seems to be isotropic within the kagome plane, while it acquires a strong out-of-plane anisotropy.
At the present stage it is difficult to establish a direct relation between the in-plane isotropic internal fields and the symmetry of the CDW. Further theoretical and experimental studies, including the exact determination of the muon-stopping site(s), are required.

\section{Conclusion}\label{seq:Conclusions}

In conclusion, a combination of zero-field and high transverse-field muon-spin rotation/relaxation experiments were performed on the CsV$_3$Sb$_5$ representative of the kagome superconducting family $A$V$_3$Sb$_5$ ($A =$K,~Cs,~Rb). The in-plane and out-of-plane electronic responses as a function of temperature and magnetic field in the normal state were studied.
An enhancement of the width of the internal magnetic field distribution sensed by the muon-spin ensemble was found to coincide with the onset of the charge ordering transition, thus suggesting that the CDW order breaks time-reversal symmetry.  A magnetic field of 8~T applied along the crystallographic $c-$axis further promotes the electronic response below $T_{\rm CDW}$, leading to a more clearly pronounced two-step increase of the internal field width at the characteristic onset temperatures $T_{\rm TRSB}=T_{\rm CDW}\simeq 95$~K and $T^\ast\simeq 30$~K, respectively.
The local fields at the muon stopping site, which are potentially created by loop currents, were found to be confined within the crystallographic $ab$-plane for temperatures between $T_{\rm CDW}$ and $T^\ast$, while they possesses a pronounced out-of-plane component below $T^\ast$.
Rotation of the crystals around the $c-$axis suggests that the internal field remains isotropic within the kagome plane, in sharp contrast to the highly anisotropic out-of-plane behaviour.
Our results indicate a rich electronic response promoted by complex charge order realized in the kagome superconductor CsV$_3$Sb$_5$ and provide useful insights into the nature of the time-reversal symmetry-breaking charge density wave order.

\begin{acknowledgments}
The work was performed at the Swiss Muon Source (S$\mu$S), Paul Scherrer Institute (PSI, Switzerland). The work of R.G. was supported by the Swiss National Science Foundation (SNF Grant No. 200021-175935). RMF was supported by the Air Force Office of Scientific Research under award number FA9550-21-1-0423. T.B. and E.R. were supported by the NSF CAREER grant DMR-2046020.
 \label{sec:Acknowledgement}
\end{acknowledgments}

\end{document}


\title{Supplementary Information for: ''Charge order breaks time-reversal symmetry in CsV$_3$Sb$_5$``}

\author{Rustem Khasanov}
 \email{rustem.khasanov@psi.ch}
 \affiliation{Laboratory for Muon Spin Spectroscopy, Paul Scherrer Institute, CH-5232 Villigen PSI, Switzerland}

\author{Debarchan Das}
 \affiliation{Laboratory for Muon Spin Spectroscopy, Paul Scherrer Institute, CH-5232 Villigen PSI, Switzerland}

\author{Ritu Gupta}
 \affiliation{Laboratory for Muon Spin Spectroscopy, Paul Scherrer Institute, CH-5232 Villigen PSI, Switzerland}

\author{Charles Mielke III}
 \affiliation{Laboratory for Muon Spin Spectroscopy, Paul Scherrer Institute, CH-5232 Villigen PSI, Switzerland}

\author{Matthias Elender}
 \affiliation{Laboratory for Muon Spin Spectroscopy, Paul Scherrer Institute, CH-5232 Villigen PSI, Switzerland}

\author{Qiangwei Yin}
 \affiliation{Department of Physics and Beijing Key Laboratory of Opto-electronic Functional Materials \& Micro-nano Devices, Renmin University of China, Beijing 100872, China}

\author{Zhijun Tu}
 \affiliation{Department of Physics and Beijing Key Laboratory of Opto-electronic Functional Materials \& Micro-nano Devices, Renmin University of China, Beijing 100872, China}

\author{Chunsheng Gong}
 \affiliation{Department of Physics and Beijing Key Laboratory of Opto-electronic Functional Materials \& Micro-nano Devices, Renmin University of China, Beijing 100872, China}

\author{Hechang Lei}
 \email{hlei@ruc.edu.cn}
 \affiliation{Department of Physics and Beijing Key Laboratory of Opto-electronic Functional Materials \& Micro-nano Devices, Renmin University of China, Beijing 100872, China}

\author{Ethan Ritz}
 \affiliation{Department of Chemical Engineering and Materials Science, University of Minnesota, MN 55455, USA}

\author{Rafael M. Fernandes}
 \affiliation{School of Physics and Astronomy, University of Minnesota, Minneapolis, MN 55455, USA}

\author{Turan Birol}
 \affiliation{Department of Chemical Engineering and Materials Science, University of Minnesota, MN 55455, USA}

\author{Zurab Guguchia}
 \email{zurab.guguchia@psi.ch}
 \affiliation{Laboratory for Muon Spin Spectroscopy, Paul Scherrer Institute, CH-5232 Villigen PSI, Switzerland}

\author{Hubertus Luetkens}
 \email{hubertus.luetkens@psi.ch}
 \affiliation{Laboratory for Muon Spin Spectroscopy, Paul Scherrer Institute, CH-5232 Villigen PSI, Switzerland}

\maketitle

\section{${\bf CsV}_3{\bf Sb}_5$ single crystal samples for $\mu$SR experiments}

\begin{figure}[htb]
\centering
\includegraphics[width=0.8\linewidth]{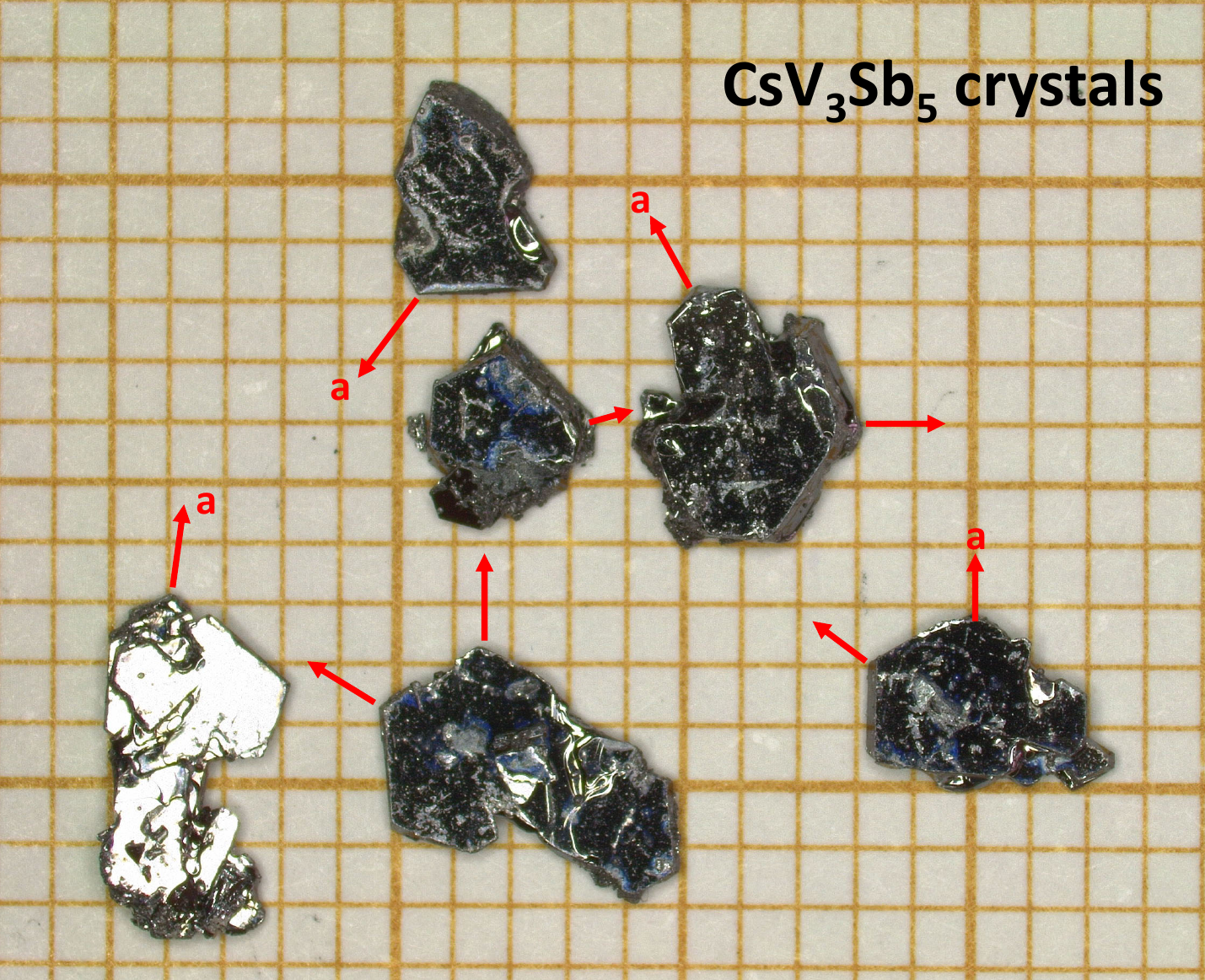}
\caption{CsV$_3$Sb$_5$ single-crystals used in muon-spin rotation/relaxation ($\mu$SR) experiments. The crystallographic $c-$axis stays perpendicularly to the flat surfaces of the crystals. The red arrows represent directions of the crystallographic $a-$axis.}
\label{fig:CsV3Sb5_crystals}
\end{figure}

Figure~\ref{fig:CsV3Sb5_crystals} shows six single crystals of CsV$_3$Sb$_5$. The crystallographic $c-$axis is perpendicular to the flat surfaces of the crystals.  The hexagonal symmetry of the crystals is obviously connected to the crystal shape. The red arrows represent directions of the crystallographic $a-$axis.

\begin{figure}[htb]
\centering
\includegraphics[width=1\linewidth]{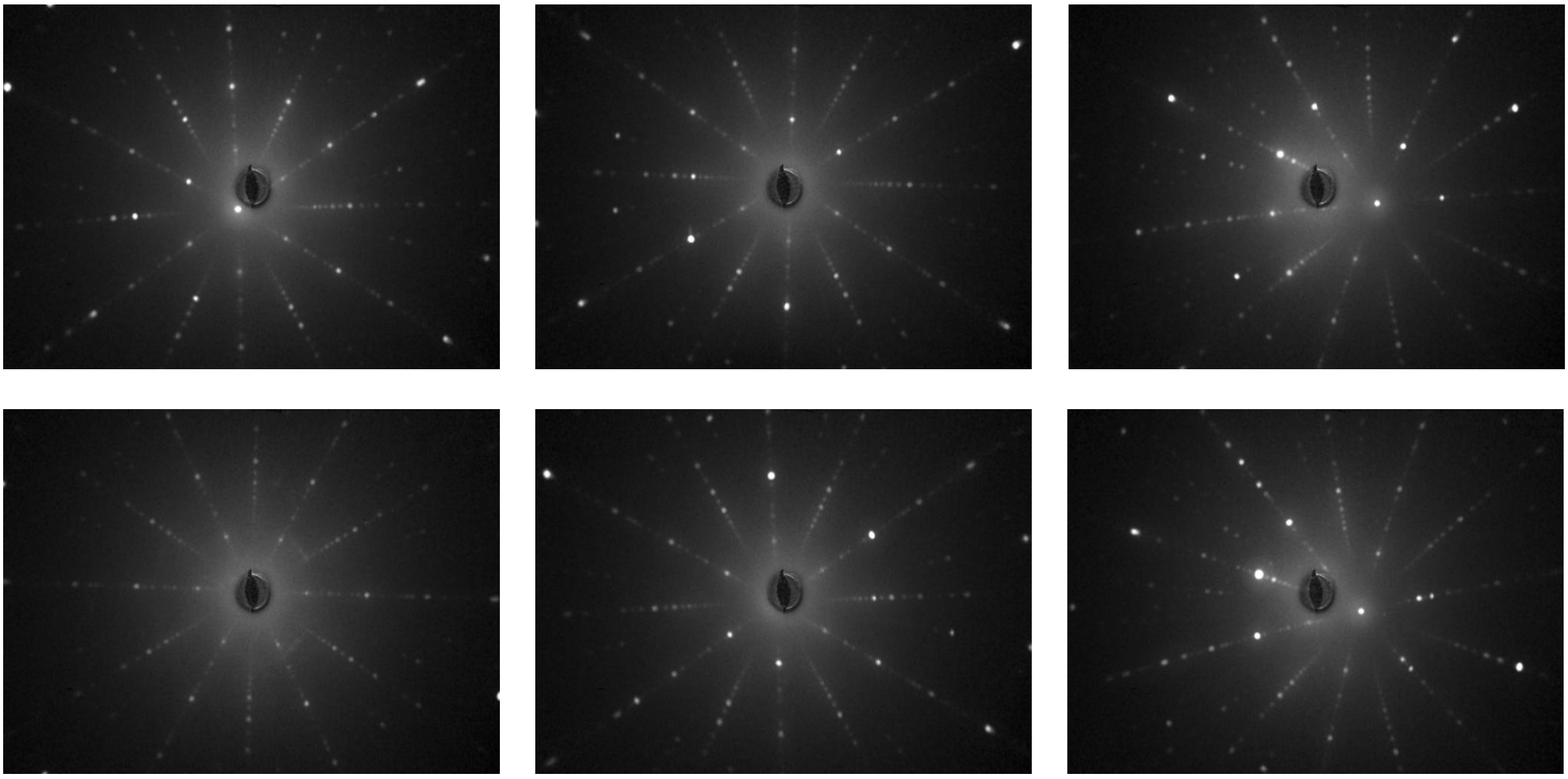}
\caption{The Laue x-ray images of the six CsV$_3$Sb$_5$ single-crystals presented in Fig.~\ref{fig:CsV3Sb5_crystals}. The images demonstrate the single crystallinity of the material and correspondence of the main crystal axes to the sample shape.}
\label{fig:Laue_CsV3Sb5_crystals}
\end{figure}

The Laue x-ray images of  six CsV$_3$Sb$_5$ crystals from Fig.~\ref{fig:CsV3Sb5_crystals} are presented in Fig.~\ref{fig:Laue_CsV3Sb5_crystals}. All images clearly demonstrate the hexagonal in-plane crystal structure of CsV$_3$Sb$_5$. The single crystallinity of the material and the correspondence of the main crystal axes to the sample shape are visible.

\pagebreak

\section{ac susceptibility response of ${\bf CsV}_3{\bf Sb}_5$ single-crystals}

\begin{figure}[htb]
\centering
\includegraphics[width=0.8\linewidth]{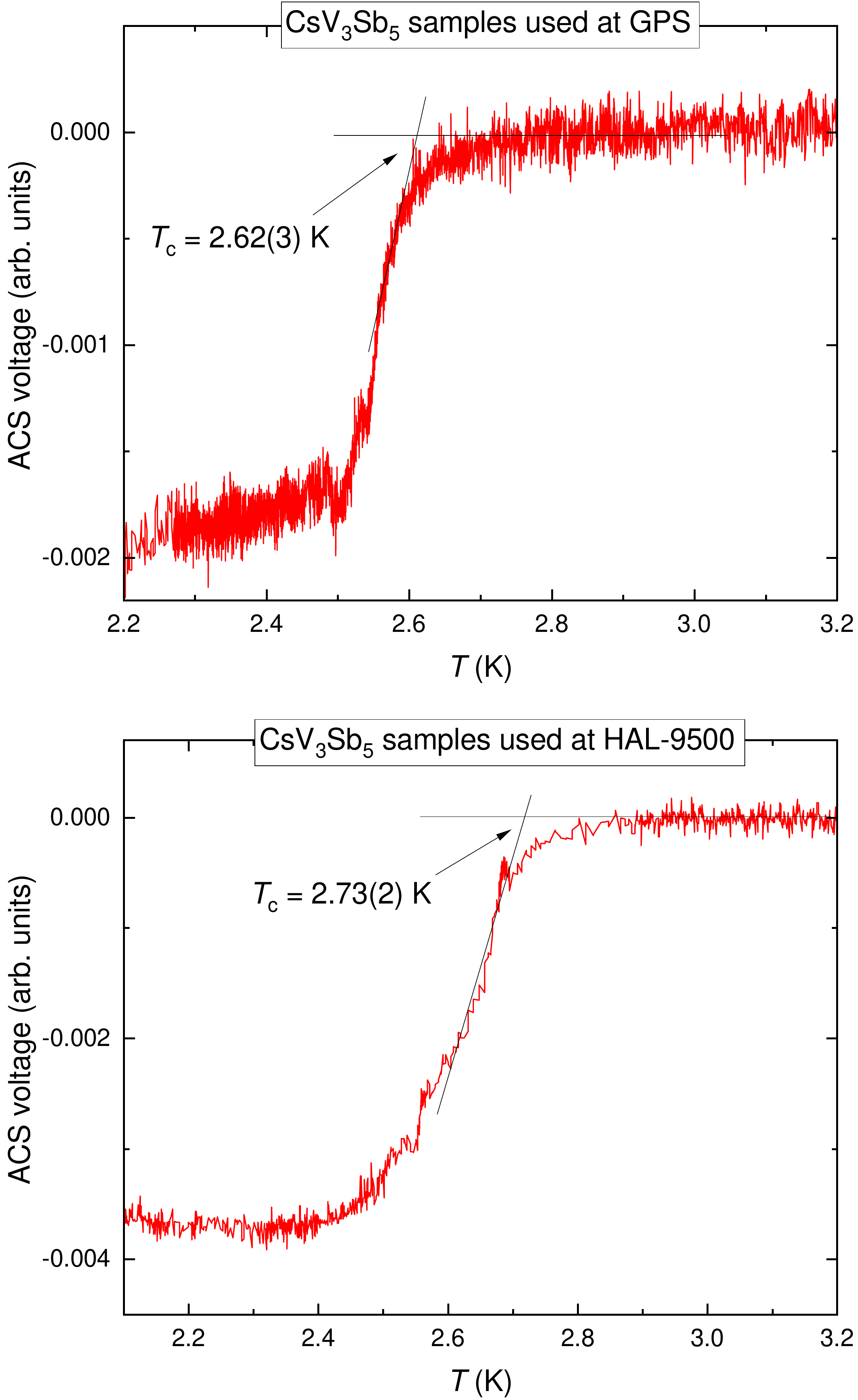}
\caption{The temperature dependence of the ac susceptibility of CsV$_3$Sb$_5$ single-crystals used in ZF-$\mu$SR [panel (a)] and TF-$\mu$SR [panel (b)] experiments, as performed at the GPS and HAL-9500 spectrometers, respectively. The transition temperature $T_{\rm c}$ is obtained from the crossing point of the linearly extrapolated ACS curves in the vicinity of $T_{\rm c}$. }
\label{fig:CsV3Sb5_ACS}
\end{figure}

The superconducting response of CsV$_3$Sb$_5$ crystals was studied via ac susceptibility (ACS) experiments. CsV$_3$Sb$_5$ single crystals were placed inside the opened pressure cell (in our ACS experiments the pressure cell was not sealed and it was simply used a sample container). The ACS setup and the pressure cell are described in Refs.~\onlinecite{Khasanov_HPR_2016, Shermadini_HPR_2017}.

Figures~\ref{fig:CsV3Sb5_ACS} shows the ACS response of CsV$_3$Sb$_5$ single-crystals used in ZF-$\mu$SR [panel (a)] and TF-$\mu$SR [panel (b)] experiments as performed at GPS and HAL-9500 spectrometers, respectively. The transition temperatures $T_{\rm c}\simeq 2.62(3)$ and $T_{\rm c}\simeq 2.72(3)$ are determined from the crossing point of the linear fits of the ACS curve above and below $T_{\rm c}$.

\begin{figure}[htb]
\centering
\includegraphics[width=1\linewidth]{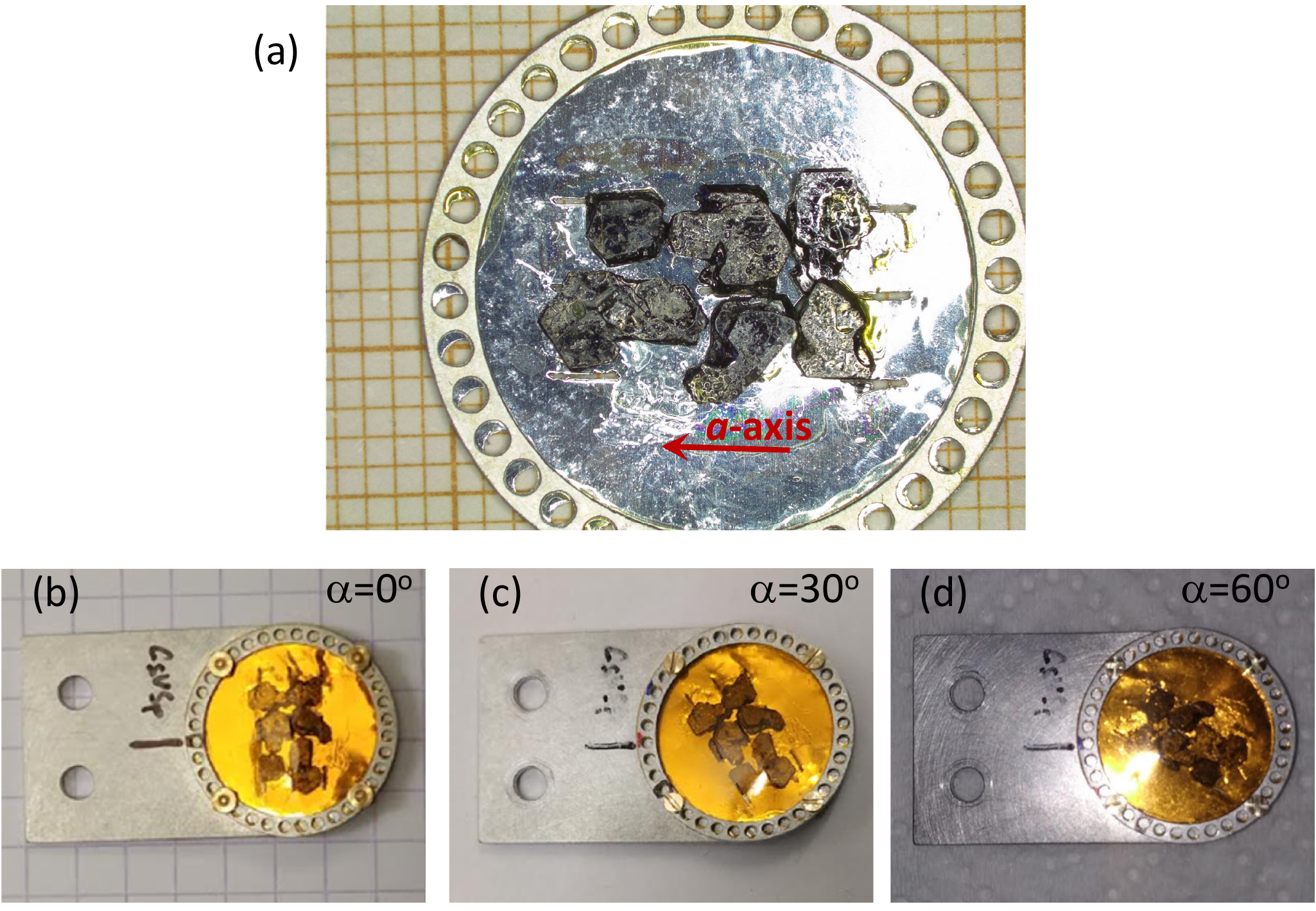}
\caption{(a) CsV$_3$Sb$_5$ single-crystals mounted on the holding ring of the GPS sample holder. (b), (c), and (d) the sample ring attached to the sample holder and turned at the angle $\alpha=0$ [panel (a)], 30 [panel (b)] and $60^{\rm o}$ [panel (c)], respectively. }
\label{fig:sample-holder-GPS}
\end{figure}

\begin{figure}[htb]
\centering
\includegraphics[width=1\linewidth]{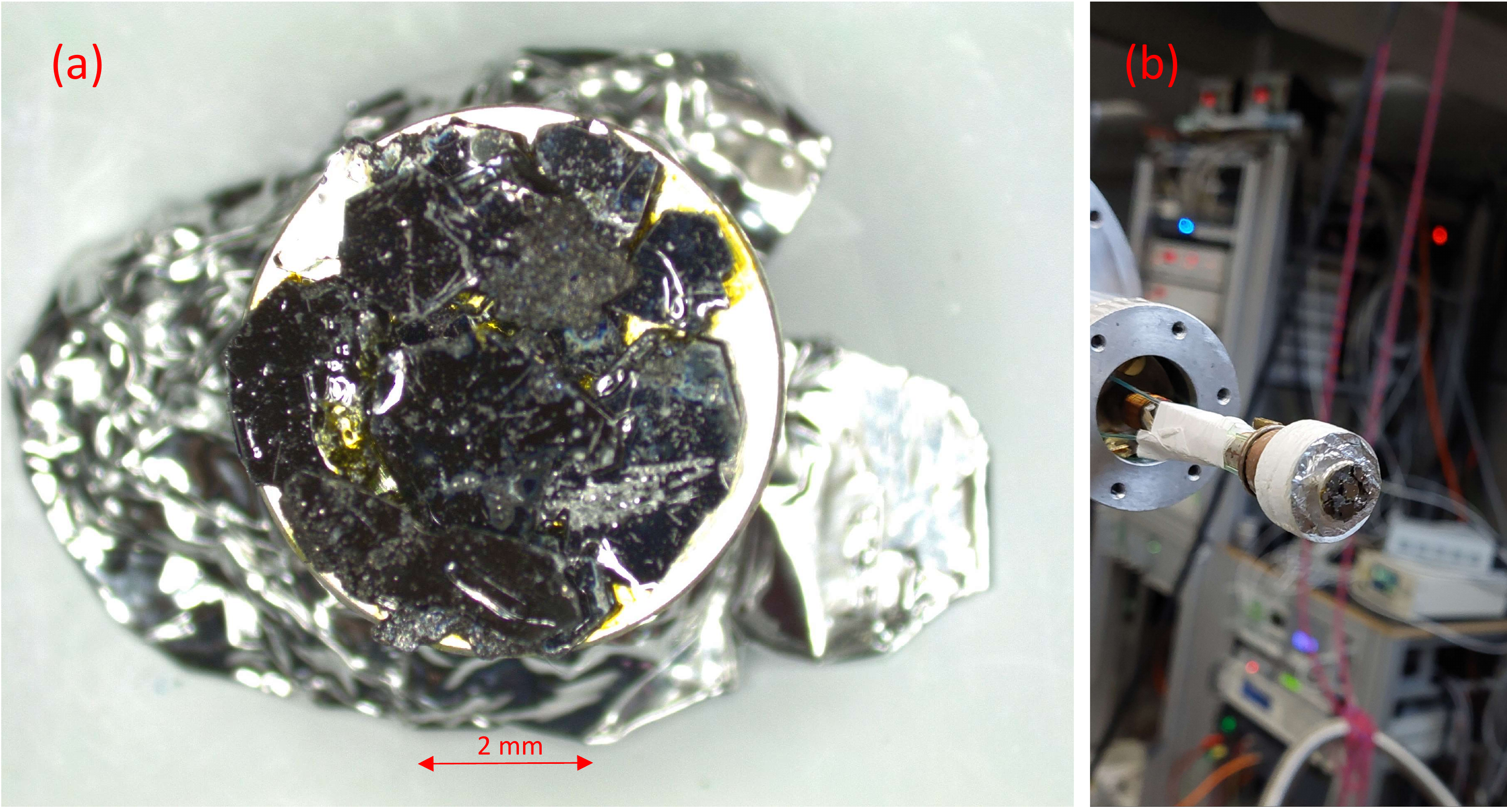}
\caption{(a) CsV$_3$Sb$_5$ single-crystals mounted on the HAL-9500 sample holder. (b) The sample holder with crystals attached to the cold finger of the continuous-flow He4 cryostat. }
\label{fig:sample-holder-HAL}
\end{figure}

\section{CsV$_3$Sb$_5$ single-crystals on $\mu$SR sample holders}

Figure~\ref{fig:sample-holder-GPS} illustrates the mounting of CsV$_3$Sb$_5$ single-crystals on the GPS sample holder. Panel (a) shows the $a-$ and $c-$axis aligned CsV$_3$Sb$_5$ crystals attached to the sample mounting ring. The ring has 36 holes allowing to rotate it with a $10^{\rm o}$ step relative to the support plate. Panels (b), (c), and (d) correspond to the positioning of the samples at angle $\alpha=0^{\rm o}$, $30^{\rm o}$, and $60^{\rm o}$, respectively.

Figure~\ref{fig:sample-holder-GPS}~(a) shows CsV$_3$Sb$_5$ single-crystals mounted on top of the HAL-9500 sample holder (made of 99.999\% pure silver). Panel (b) demonstrates the sample holder attached to the cold finger of the continuous-flow He4 cryostat.

\section{The background estimate}

In $\mu$SR experiments, part of the muons may not hit the sample and stop in the sample holder, cryostat windows, cryostat walls, {\it etc.} These muons contribute in the background response, which must be known prior to performing the data analysis.

\subsection{The background contribution in GPS experiments}

In experiments performed at the GPS spectrometer, the samples (CsV$_3$Sb$_5$ single-crystals) were mounted on the holder shown in Fig.~\ref{fig:sample-holder-GPS}. The background was estimated from measurements made in the superconducting state. The external magnetic field $B_{\rm ext}=10$~mT was applied parallel to the crystallographic $c-$axis at $T=5$~K, {\it i.e.} above the superconducting transition temperature $T_{\rm c}\simeq 2.6$~K [see Fig.~\ref{fig:CsV3Sb5_ACS}~(a)]. The angles $\alpha$ and $\beta$ {\it i.e.} the angle between the crystallographic $c-$axis and the in-plane component of the muon-spin polarization ${\bf P}^{\parallel ab}_\mu$ and the angle between the muon momentum ${\bf p}_\mu$ and the initial muon-spin polarization ${\bf P}_\mu$, we kept at $\alpha=30^{\rm o}$ and $\beta=45^{\rm o}$, respectively.

\begin{figure}[htb]
\centering
\includegraphics[width=0.8\linewidth]{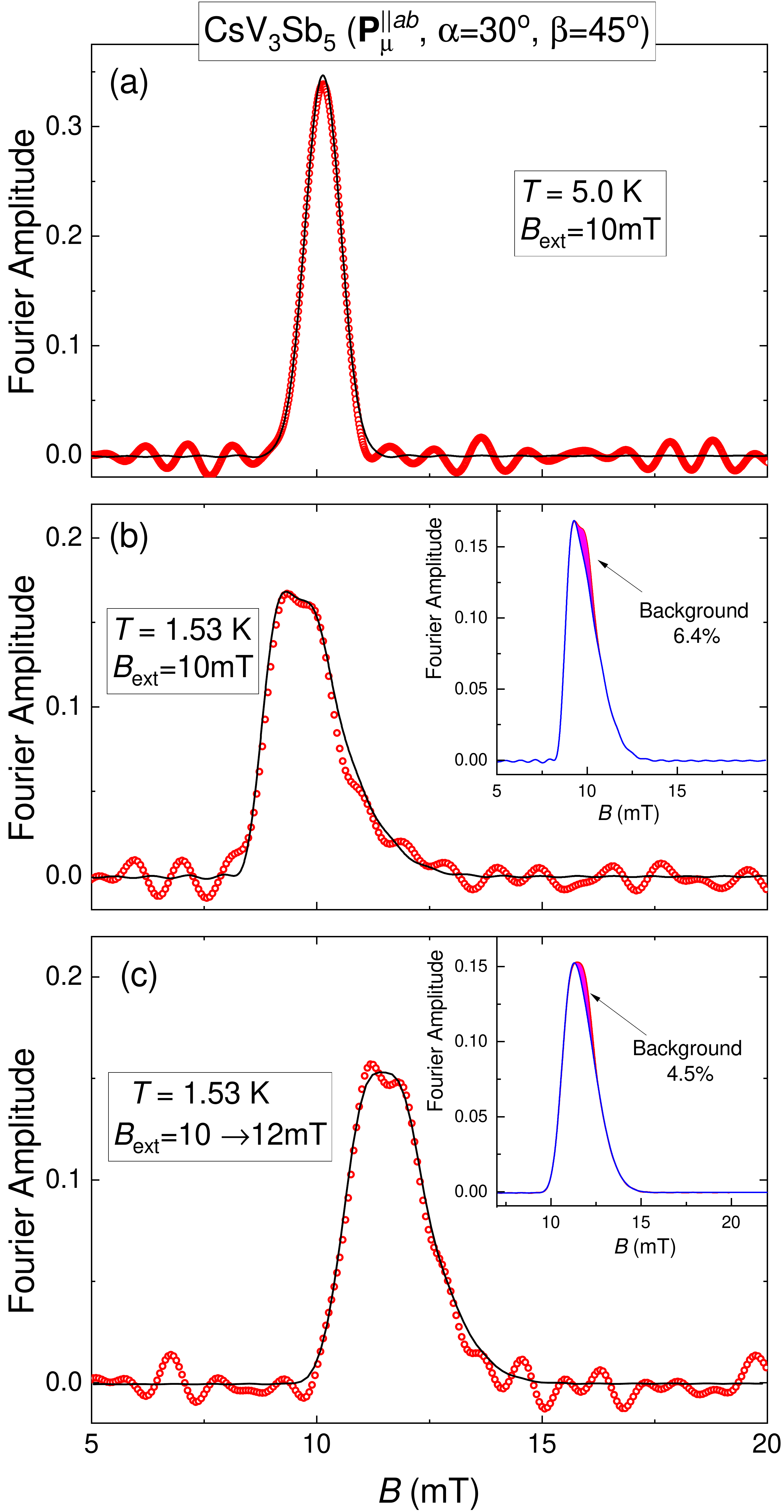}
\caption{(a) The magnetic field distribution $P(B)$ in CsV$_3$Sb$_5$ single-crystals mounted on the GPS sample holder and measured in the normal state at $B_{\rm ext}=10$~mT and $T=5.0$~K. (b) The magnetic field distribution in CsV$_3$Sb$_5$ single-crystals after field cooling in $B_{\rm ext}=10$~mT down to $T\simeq 1.53$~K. The asymmetric line shape of $P(B)$ is caused by formation of the flux-line lattice (FLL). The inset compares the 'total` $P(B)$ with the contribution of the FLL. The background contribution is marked in pink. (c) The same as panel (b), but after shifting the external field $B_{\rm ext}$ from 10 to 12~mT.  }
 \label{fig:Fourier-GPS}
\end{figure}

The Fourier transforms of TF-$\mu$SR time spectra, representing the internal field distribution $P(B)$, are shown in Fig.~\ref{fig:Fourier-GPS}. The panel (a) corresponds to $P(B)$ obtained at $T=5$~K. The other two panels represent internal field distributions in the superconducting state, {\it i.e. }after cooling the sample down to $T\simeq1.53$~K at $B_{\rm ext}=10$~mT [panel (a)] and after subsequent increase of $B_{\rm ext}$ from 10 to 12~mT, which occurs without changing temperature [panel (c)]. The red curves correspond to the experimental data. The solid black lines are fits by assuming the separation of the TF-$\mu$SR response of the sample (S) and the background (BG) components:
\begin{equation}
A(t)=A_{0, {\rm S}} \; P_{\rm S}(t) + A_{0, {\rm BG}} \; P_{\rm BG}(t).
 \label{eq:P_TF_full}
 \end{equation}
Here $A_{0, {\rm S}}$ and $A_{0, {\rm BG}}$ are the initial asymmetries, while $P_{\rm S}(t)$ and $P_{\rm BG}(t)$ are the time evolution of the muon-spin polarizations of the sample and the background component, respectively. The background contribution was approximated by a cosine decay function:
\begin{equation}
P_{\rm BG}(t) =  \exp(-\sigma_{\rm BG}^2 t^2/2) \cos (\gamma_\mu B_{\rm ext} t +\phi).
 \label{eq:BG_TF}
\end{equation}
Here $\gamma_\mu=2\pi 135.5$~MHz/T is the muon gyromagnetic ratio, $\sigma_{\rm BG}$ is the Gaussian relaxation rate and $\phi$ is the initial phase of the muon-spin ensemble.
In order to account for the asymmetric field distribution $P(B)$ caused by formation of the flux-line lattice (FLL) in the superconducting state, the sample contribution was described using the  Skewed Gaussian (SKG) distribution function:\cite{Suter_2008}
\begin{equation}
P_{\rm S}(t) = {\rm SKG}(t) \; \cos (\gamma_\mu B_{\rm 0} t +\phi).
 \label{eq:S_TF}
\end{equation}
Here $B_0$ is the field corresponding to the maximum of $P(B)$ distribution of ${\rm SKG}(t)$ function.\cite{Suter_2008}
The fit of Eq.~\ref{eq:P_TF_full} with the sample and the background components described by Eqs.~\ref{eq:BG_TF} and \ref{eq:S_TF} results in $A_{0,{\rm BG}}/A_0\simeq 0.064$ and 0.045 for the experimental data presented in Figs.~\ref{fig:Fourier-GPS}~(c) and (d), respectively. As an average value, $\sim5$\% background contribution was considered for experiments performed at the GPS spectrometer.

The background contribution might be visualized by comparing the 'total` $P(B)$ distributions  with the field distributions caused by formation of the FLL inside the superconducting CsV$_3$Sb$_5$ [see the corresponding red and blue lines at insets to Figs.~\ref{fig:Fourier-GPS}~(c) and (d)]. The difference, as is marked by pink areas, represent the background contribution.

\begin{figure}[htb]
\centering
\includegraphics[width=0.8\linewidth]{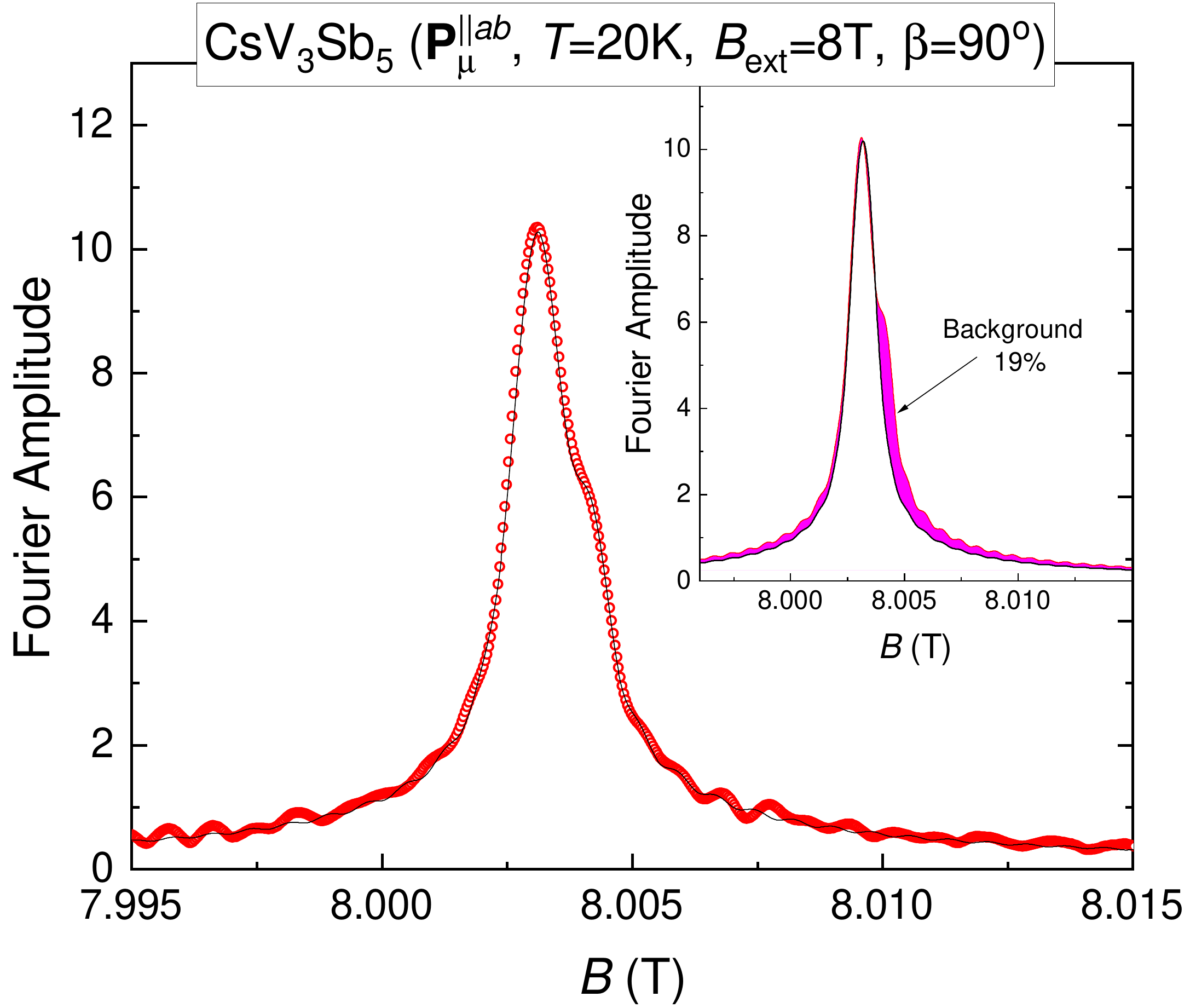}
\caption{The magnetic field distribution $P(B)$ in CsV$_3$Sb$_5$ single-crystals mounted on the HAL-9500 sample holder and measured at $B_{\rm ext}=8.0$~T and $T=20.0$~K. The inset compares the 'total` $P(B)$ with the contribution of the sample. The background contribution is marked in pink. }
 \label{fig:Fourier-HAL}
\end{figure}

\begin{figure*}[htb]
\centering
\includegraphics[width=1.0\linewidth]{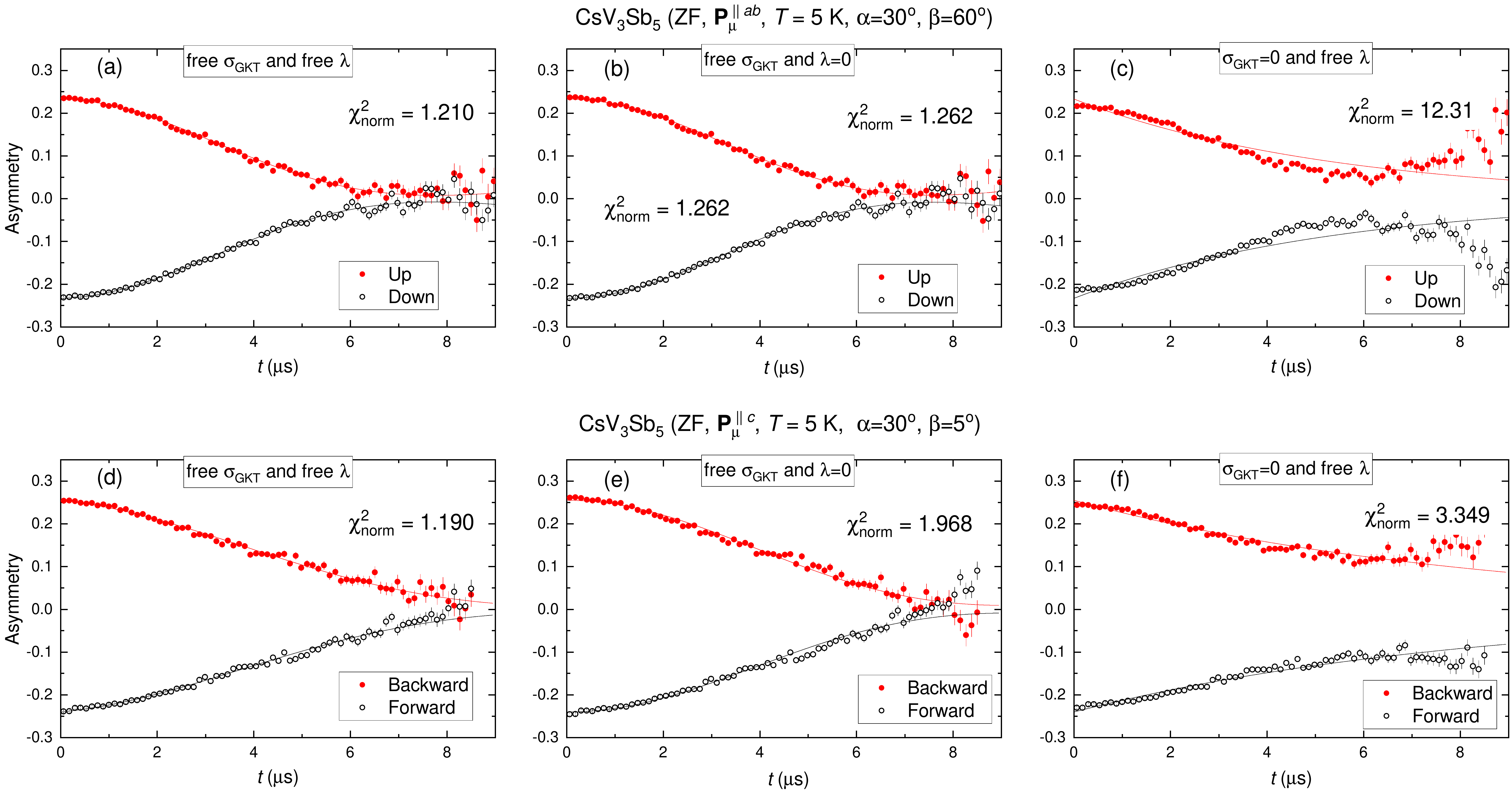}
\caption{(a)--(c) The asymmetry spectra collected for the $P^{\parallel ab}(t)$ set of experiments at $T=5$~K, $\alpha=30^{\rm o}$ and $\beta=60^{\rm o}$. The red and black curves correspond to the response of positron detectors staying in-phase and out-of-phase with ${\bf P}^{\parallel ab}$. (d)--(f) The asymmetry spectra collected for the $P^{\parallel ab}(t)$ set of experiments at $T=5$~K, $\alpha=30^{\rm o}$ and $\beta=5^{\rm o}$. The solid lines in panels (a)--(f) are fits of Eq.~1 from the main text to the experimental data. The fits were made with both -- the Gaussian Kubo-Toyabe $\sigma_{\rm GKT}$ and the exponential $\lambda$ -- relaxations remaining free [panels (a) and (d)]; with $\sigma_{\rm GKT}$ and $\lambda$ fixed to zero [panels (b) and (e)]; and with $\sigma_{\rm GKT}=0$ and $\lambda$ free [panels (c) and (f)]. }
\label{fig:time-spectra-5K}
\end{figure*}

\subsection{Background for experiments performed at the HAL-9500 spectrometer}

In high-field experiments performed at the HAL-9500 spectrometer under an applied field of $B_{\rm ext}=8$~T, a two-component signal was observed within the full range of temperatures. Figure~\ref{fig:Fourier-HAL} shows the field distribution $P(B)$ measured at $T=20$~K. The solid line corresponds to the two component fit by means of Eq.~\ref{eq:P_TF_full}. The background and the sample component were described by the aforementioned Eq.~\ref{eq:BG_TF} and by Eq.~2 from the main text, respectively. The background contribution is visualized as a difference between the two-component fit and the sample contribution (see the pink area at the inset to Fig.~\ref{fig:Fourier-HAL}).

The width and the position of the background signal were found to be temperature independent, as expected, and they were kept fixed during the analysis ($A_{0,{\rm BG}}/A_0\simeq 0.2$, $\sigma_{\rm BG}\simeq 0.16$~$\mu {\rm s}^{-1}$).

\section{The presence of nuclear and electronic components in $\mu$SR response  }\label{sec:Data_Analysis}

Validity of the data analysis procedure considering the presence of both -- the electronic and the nuclear components -- was checked for the set of ZF-$\mu$SR data collected at $T=5$~K for two different components of the muon-spin polarization [$P^{\parallel ab}(t)$ and $P^{\parallel c}(t)$]. The red and black curves correspond to the data collected at positron detectors staying in-phase ($0^{\rm o}$) and out-of-phase ($180^{\rm o}$) to the corresponding initial component of the muon-spin polarization. Three different fit types of Eq.~1 from the main text to the experimental data were performed:  \\
1. Both, $\lambda$ and $\sigma_{\rm GKT}$, remain free, {\it i.e.}, stay fitted [panels (a) and (d)]\\
2. $\sigma_{\rm GKT}$ stays free and $\lambda$ is fixed to `0' [panels (b) and (e)].\\
3. $\lambda$ remains free and $\sigma_{\rm GKT}=0$ [panels (c) and (f)].\\
The results of the fit of Eq.~1 from the main text to the data are presented in Fig.~\ref{fig:time-spectra-5K} by solid lines.

\begin{table}[htb]
     \centering
     \caption{The results of the fit of Eq.~1 from the main text to the muon data recorded at $T=5$~K for two different components of the muon-spin polarization $P^{\parallel ab}(t)$ and $P^{\parallel c}(t)$.  \\ }
     \begin{tabular}{c|ccccc}
 \hline  \hline
    & Fit type & {\centering $\lambda$} & $\sigma_{\rm GKT}$ & $\chi^2_{\rm norm}$ \\
        &          &($\mu{\rm s}^{-1}$)&($\mu{\rm s}^{-1}$ )&\\
 \hline
        &free $\lambda$ and $\sigma_{\rm GKT}$  & 0.0118(40) & 0.2231(22) &	1.210  \\
${\bf P}^{\parallel ab}_\mu$ &$\lambda=0$ and free $\sigma_{\rm GKT}$& 0 & 0.2281(13) &	1.262  \\
        &free $\lambda$ and $\sigma_{\rm GKT}=0$& 0.1857(25) & 0 &	12.31  \\
 \hline
                &free $\lambda$ and $\sigma_{\rm GKT}$  & 0.0453(47)\ \  & 0.1593(42) \ \ & 1.190 \\
${\bf P}^{\parallel c}_\mu$&\ $\lambda=0$ and free $\sigma_{\rm GKT}$ \ & 0 & 0.1900(16)& 1.968 \\
                &free $\lambda$ and $\sigma_{\rm GKT}=0$& 0.1204(25) & 0 & 3.349 \\
 \hline  \hline
     \end{tabular}
     \label{tab:Fit-results}
 \end{table}

The fitted relaxation rates are summarised in Table~\ref{tab:Fit-results}. The goodness of each fit type was evaluated by calculating the values of the normalized $\chi^{2}_{\rm norm}$ (the sum of the mean squared deviations divided by the number of degrees of freedom). Note that, in their normalized form, $\chi^{2}_{\rm norm}$ takes into account the decreased number of the fit parameters for cases when one of the relaxation rates ($\lambda$ or $\sigma_{\rm GKT}$) was fixed to zero.

The results presented in Fig.~\ref{fig:time-spectra-5K} and Table~\ref{tab:Fit-results} imply that the experimental data require the {\it presence of both} -- the exponential $\lambda$ and the Gaussian Kubo-Toyabe $\sigma_{\rm GKT}$ -- relaxation components. Exclusion of either $\lambda$ or $\sigma_{\rm GKT}$ leads to substantial increase of $\chi^2_{\rm norm}$.
This agrees, therefore, with the results of Ref.~\onlinecite{Mielke_arxiv_2021}, where the ZF-$\mu$SR data of KV$_3$Sb$_5$ representative of $A$V$_3$Sb$_5$ family were described by using both -- $\sigma_{\rm GKT}$ and $\lambda$ relaxation components. A similar approach was used in the majority of other TRSB $\mu$SR studies as reported {\it e.g. } in Refs.~\onlinecite{Luke_PRL_1993, Luke_Nature_1998, Maisuradze_PRB_2010, Hillier_PRL_2012, Biswas_PRB_2013, Shang_PRL_2018, Ghosh_JPCM_2021, Grinenko_Ba122_NatPhys_2020, Grinenko_SRO_NatPhys_2020, Grinenko_SRO_NatCom_2021}.
Our results would also suggest that the analysis of CsV$_3$Sb$_5$ and KV$_3$Sb$_5$ ZF-$\mu$SR data from Refs.~\onlinecite{Yu_arxiv_2021} and \onlinecite{Kenney_JPCM_2020}, where only the Gaussian Kubo-Toyabe term was taken into account, need to be reconsidered.

\section{Comparison between the zero-field and longitudinal-field data  }\label{sec:ZF-LF}

\begin{figure}[htb]
\centering
\includegraphics[width=0.8\linewidth]{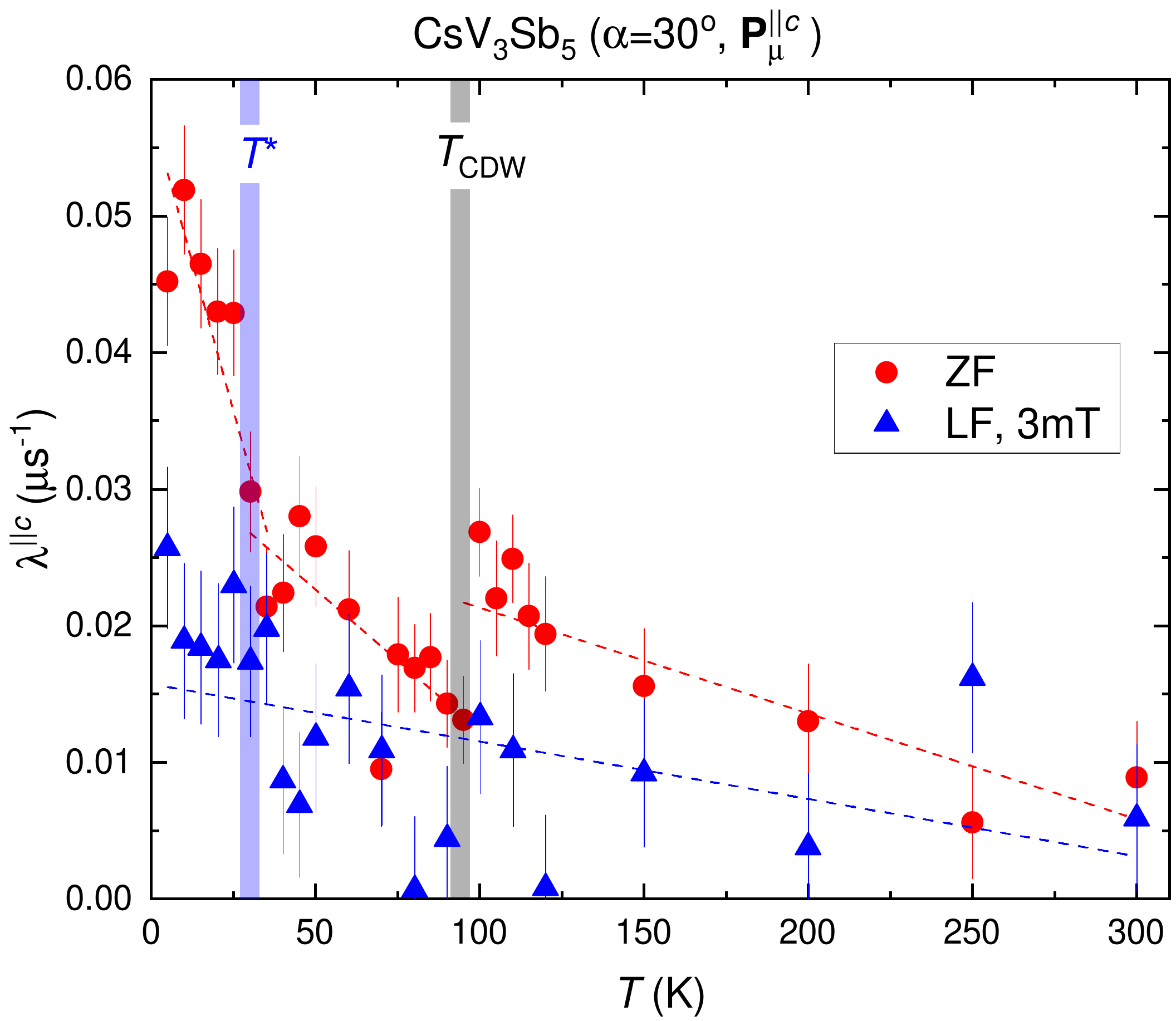}
\caption{The exponential relaxation rates measured in zero-field and longitudinal-field (LF) $\mu$SR experiments. In LF-$\mu$SR experients the external magnetic field of $B_{\rm ext}=3$~mT was applied along the ${\bf P}_\mu^{\parallel ab}$ component of the muon-spin polarization [see Fig.~1(a) in the main text]. The dashed lines are the linear fits to the ZF and LF $\lambda^{\parallel c}(T)$ data.}
 \label{fig:LF}
\end{figure}

Both static and fluctuating internal magnetic fields can have an effect on the exponential muon-spin polarization relaxation rate as is measured in ZF experiments. In order to differentiate between static and fluctuating internal magnetic fields, the longitudinal-field (LF) experiments with the externally applied magnetic field $B_{\rm ext}=3$~mT applied along the ${\bf P}_\mu^{\parallel ab}$ component of the muon-spin polarization were conducted. The results reveal that the Gaussian Kubo-Toyabe component observed in ZF-$\mu$SR is absent, as expected when the muon-spins are decoupled from the local field, while the exponential relaxation rate $\lambda$ remains at a certain non-zero value.

Figure~\ref{fig:LF} compares the exponential relaxation rates as they observed in ZF and LF set of $\mu$SR experiments. From the data presented in Fig.~\ref{fig:LF} two important points emerge:\\
(i) The LF relaxation points stay systematically lower the ZF ones. This suggests that 3~mT magnetic field leads to partial, but not complete, decoupling of muon-spins from internal fields. \\
(ii) The LF relaxation rate decreases nearly linearly within the full temperature range studied, {\it i.e.} from $\simeq 3$ up to $\simeq 300$~K. This imply that the dynamical component on the exponential relaxation is indeed present. However it is not related to the appearance of the charge density order below $T_{\rm CDW}\simeq 95$~K, as well as to an additional enhancement of the ZF relaxation below $T^\ast\simeq30$~K.